%% file: Manuscript.tex
\documentclass[iicol, sn-mathphys-num]{sn-jnl}% Charles
\usepackage{graphicx}%
\usepackage{multirow}%
\usepackage{amsmath,amssymb,amsfonts}%
\usepackage{amsthm}%
\usepackage{mathrsfs}%
\usepackage[title]{appendix}%
\usepackage{xcolor}%
\usepackage{textcomp}%
\usepackage{manyfoot}%
\usepackage{booktabs}%
\usepackage{algorithm}%
\usepackage{algorithmicx}%
\usepackage{algpseudocode}%
\usepackage{listings}%
\usepackage{tikz,pgf}
\usepackage{pgfplots}
\pgfplotsset{compat=1.16}
\usetikzlibrary{shapes}
\usetikzlibrary{arrows}
\usetikzlibrary{plotmarks,spy,calc,patterns,fit}

\usepackage{nicefrac}
\usepackage{bm} % bold face for math

\usepackage{hyperref}

\usepackage{caption}
\usepackage{subcaption}

\definecolor{MyBlue}{rgb}{0.00000,0.44700,0.74100}%
\definecolor{MyDarkGreen}{HTML}{006100}
\definecolor{MyMagenta}{rgb}{0.6, 0.4, 0.8}
\definecolor{MyOrange}{rgb}{0.93, 0.57, 0.13}

\newcommand{\plotfigureheight}{0.72}

\usepackage[acronym]{glossaries}

\hypersetup{%
  colorlinks=true,  
  linkcolor=black, 
  linktoc=all,
}

\theoremstyle{thmstyleone}%
\theoremstyle{thmstyletwo}%

\theoremstyle{thmstylethree}%
\newtheorem{definition}{Definition}%

\newacronym{bpsk}{BPSK}{binary phase-shift keying}
\newacronym{awgn}{AWGN}{additive white Gaussian noise}
\newacronym{5g}{5G}{3GPP's next-generation mobile-communication standard}
\newacronym{snr}{SNR}{signal-to-noise ratio}
\newacronym{fer}{FER}{frame-error rate}
\newacronym{ldpc}{LDPC}{low-density parity check}
\newacronym{sc}{SC}{successive-cancellation}
\newacronym{scl}{SCL}{successive-cancellation list}
\newacronym{scf}{SCF}{successive-cancellation flip}
\newacronym{dscf}{DSCF}{dynamic SCF}
\newacronym{pscf}{PSCF}{partitioned SCF}
\newacronym{pbf}{PBF}{progressive bit-flipping}
\newacronym{bp}{BP}{belief propagation}

\newacronym{bpf}{BPF}{belief propagation flip}

\newacronym{scan}{SCAN}{soft cancellation}

\newacronym{frm}{FRM}{flexible restart mechanism}

\newacronym{crc}{CRC}{cyclic-redundancy check}
\newacronym{llr}{LLR}{log-likelihood ratio}
\newacronym{ps}{PS}{partial-sum}
\newacronym{cc}{CC}{clock cycle}
\newacronym{srm}{SRM}{simplified restart mechanism}
\newacronym{grm}{GRM}{generalized restart mechanism}
\newacronym{llrm}{LLRM}{limited-location restart mechanism}
\newacronym{rhs}{RHS}{right-hand side}
\newacronym{lhs}{LHS}{left-hand side}
\newacronym{pdf}{PDF}{probability-density function}
\newacronym{pmf}{PMF}{probability-mass function}
\newacronym{mmtc}{mMTC}{massive machine-type communications}
\newacronym{embb}{eMBB}{enhanced mobile broadband}
\newacronym{lsb}{LSB}{least significant bit}
\newacronym{urllc}{URLLC}{ultra-reliable low-latency communications}
\newacronym{ssc}{SSC}{Simplified SC}
\newacronym{spc}{SPC}{single-parity-check}
\newacronym{rep}{REP}{repetition}
\newacronym{lrt}{LRT}{latency-reducing technique}
\newacronym{rone}{R1}{rate-1}
\newacronym{rzero}{R0}{rate-0}
\newcommand{\scLRT}{\text{SC}_{\text{LRT}}} %sc implemented in PolarBear
\newcommand{\scmodif}{$\text{SC}\left(\psi_{t'}, \bm{\varepsilon}_{t'} \right)$\,}
\newcommand{\scmodifmath}{\text{SC}\left(\psi_{t'}, \bm{\varepsilon}_{t'} \right)}
\newcommand{\fssc}{FSSC}
\newcommand{\fsscmodif}{$\text{FSSC}\left(\psi_{t'}, \bm{\varepsilon}_{t'} \right)$\,}
\newcommand{\fsscbitflip}{$\text{FSSC}\left(\bm{\varepsilon}_{t'} \right)$\,}
\newcommand{\dec}{\text{dec}}

\newcommand{\flip}{\text{flip}}
\newcommand{\flipgrm}{\text{flipGRM}}

\newcommand{\avgreducgrm}{\Delta\overline{\mathcal{L}}_{\flip}}

\newcommand{\HD}{\text{HD}}

\newcommand{\Tmax}{T_{\text{max}}}

\begin{document}

\title[Generalized Restart Mechanism for Successive-Cancellation Flip Decoding of Polar Codes]{Generalized Restart Mechanism for Successive-Cancellation Flip Decoding of Polar Codes}


\author*[1]{\fnm{Ilshat} \sur{Sagitov}}\email{ilshat.sagitov.1@ens.etsmtl.ca}

\author*[1]{\fnm{Charles} \sur{Pillet}}\email{charles.pillet.1@ens.etsmtl.ca}

\author*[2]{\fnm{Alexios} \sur{Balatsoukas-Stimming}}\email{a.k.balatsoukas.stimming@tue.nl}

\author*[1]{\fnm{Pascal} \sur{Giard}}\email{pascal.giard@etsmtl.ca}

\affil*[1]{\orgdiv{LaCIME, Department of Electrical Engineering}, \orgname{\'Ecole de technologie sup\'erieure (\'ETS}, \orgaddress{\street{1100 Notre-Dame St West}, \city{Montréal}, \postcode{H3C 1K3}, \state{Québec}, \country{Canada}}}

%\affil[2]{\orgdiv{LaCIME, Department of Electrical Engineering}, \orgname{Organization}, \orgaddress{\street{Street}, \city{City}, \postcode{10587}, \state{State}, \country{Country}}}

\affil*[2]{\orgdiv{Department of Electrical Engineering}, \orgname{Eindhoven University of Technology}, \orgaddress{\street{Groene Loper 19}, \city{Eindhoven}, \postcode{5600 MB}, \country{The Netherlands}}}


%Polar codes have received a lot of attention due to their capacity-achieving property for practically relevant communication channels with low-complexity \acrfull{sc} decoding. 
%However, the error-correction performance of the  \gls{sc} algorithm is very limited at short to moderate code lengths.
\abstract{
Polar codes are a class of linear error-correction codes that have received a lot of attention due to their ability
to achieve channel capacity in an arbitrary binary discrete memoryless channel (B-DMC) with low-complexity
\acrfull{sc} decoding. However, practical implementations often require better error-correction
performance than what \gls{sc} decoding provides, particularly at short to moderate code lengths. \Acrfull{scf} decoding algorithm was proposed to improve error-correction performance with an aim to detect and correct the first wrongly estimated bit in a codeword before resuming \gls{sc} decoding. At each additional \gls{sc} decoding trial, i.e., decoding attempt beyond the initial unsuccessful trial, one bit estimated as the least reliable is flipped. \Acrfull{dscf} is a variation of \gls{scf}, where multiple bits may be flipped simultaneously per trial. Despite the improved error-correction performance compared to the \gls{sc} decoder, \gls{scf}-based decoders have variable execution time, which leads to high average execution time and latency. 
%at the cost of a variable execution time,  and a high latency (worst-case execution time). 
In this work, we propose the \acrfull{grm} that allows to skip decoding computations that are identical between the initial trial and any additional trial.
%without affecting the error-correction performance. 
Under \gls{dscf} decoding with up to 3-bit flips per decoding trial, our proposed \gls{grm} is shown to reduce the average execution time by 25\% to 60\% without any negative effect on error-correction performance. The proposed mechanism is adaptable to state-of-the-art latency-reduction techniques. When applied to Fast-\gls{dscf}-3 decoding, the additional reduction brought by the \gls{grm} is 15\% to 22\%. For the \gls{dscf}-3 decoder, the proposed mechanism requires approximately 4\% additional memory.}




\maketitle
\glsresetall
\section{Introduction}\label{sec1}

Digital communication systems are integrated into many areas of modern technology. Networks such as the Internet of Things (IoT) require low-power transmitters and receivers to provide service for networks with strict requirements on production cost and battery life. 
%Modern communication systems are in need of low-power transmitters and receivers to provide service for networks with strict requirements on battery life, such as Internet of Things (IoT) networks. 
Blocks implementing error detection and correction, particularly decoding, are typically among the most resource-intensive blocks of the entire communication system.

Polar codes are linear block codes, proposed in \cite{arik_polariz}, that were shown to asymptotically achieve the channel capacity under low-complexity \gls{sc} decoding as the code length tends to infinity.  
When using the low-complexity \gls{sc} algorithm, polar codes provide mediocre error-correction performance at short-to-moderate code lengths compared to other modern codes such as \gls{ldpc} codes \cite{pol_codes_ldpc}.
\Gls{scl} decoding \cite{scl_intro}, that keeps a list of $L$ candidate paths,
 significantly improves the error-correction performance to the extent that polar codes were selected to protect the control channel of the \gls{embb} service in \gls{5g} \cite{3GPP_5G_Coding}. Moreover, low-complexity encoding and decoding algorithms make polar codes suitable for \gls{mmtc} service of \gls{5g} networks \cite{pol_mmtc_chan,pol_ldpc_mmtc}, which fits the requirements of IoT networks. 
 
\Gls{scf} decoding was proposed in \cite{scf_intro} to improve the error-correction performance of \gls{sc} decoding.
Unlike the \gls{scl}, which decodes a codeword with $L$ parallel candidates, \gls{scf} generates the candidates across multiple decoding trials, i.e., decoding attempts.
%Unlike \gls{scl} decoding, which  uses $L$ parallel \gls{sc} instances for $L$ candidate paths, \gls{scf} decoding reuses a single \gls{sc} instance over multiple decoding trials, i.e., decoding attempts.
A list of bit-flipping candidates is constructed at the end of an initial unsuccessful \gls{sc} decoding trial. This list contains the bit locations estimated as the least reliable ones. At each additional trial, a decision bit from the list is flipped during the course of the \gls{sc} decoding before normal decoding is resumed.
%At the end of an initial unsuccessful \gls{sc} decoding trial, a list of bit-flipping candidates, which contains the bits that are estimated as the least reliable, is constructed. At each additional trial, a decision bit from the list is flipped during the course of \gls{sc} decoding before normal decoding is resumed.
%Polar-encoded words are concatenated with \gls{crc} bits that allow \gls{scf} decoding to validate an estimated word and exit decoding if the \gls{crc} passes.
 A block of information bits is concatenated with the  \gls{crc} code before passing it through a polar encoder. This allows the \gls{scf} decoder to validate an estimated information word and exit decoding when the \gls{crc} passes. 

The \gls{scf} decoding achieves the same error-correction performance as the \gls{scl} decoding with a small list size $L$ \cite{scf_intro}. To improve the error-correction performance of \gls{scf} decoding, several variations have been proposed. 
\Gls{dscf} decoding was proposed in \cite{dyn_scf} with two major contributions: (a) a more accurate metric to construct the list of the bit-flipping candidates, and (b) algorithmic modifications whereby multiple bits can be simultaneously flipped per decoding trial and where the list of candidates is dynamically updated. As a result of these modifications, the \gls{dscf} decoding achieves the same performance as the \gls{scl} with a medium list size $L$. 
In \cite{eis_scf,progr_bf_journ} improvements to \gls{scf} decoding were proposed, where the list of bit-flipping candidates is constrained to a critical set constructed offline, i.e., the most frequent erroneously-estimated bits are identified by way of simulation. The multi-bit flipping approach, \gls{pbf} decoding, was further derived in \cite{progr_bf_journ} based on a progressive update of the critical set during the course of decoding. In \cite{thr_scf}, the bit-flipping set is constructed using a predefined threshold. In \cite{multi_scf}, a multi-bit flip \gls{scf} decoding variation was derived through the offline observation of the error pattern behavior.
%Another variation of \gls{scf}, \gls{pscf} decoding was proposed by \citerefs{part_scf}. In \gls{pscf} decoding, a codeword is divided into partitions, each with its own \gls{crc}. The \gls{scf} decoding is applied to each partition separately. This technique allows to potentially exit decoding trials earlier without decoding the whole codeword. As a result, the average execution time is reduced. Also, partitioning improves the error-correction performance compared to the standard \gls{scf}. Hardware implementation of the \gls{pscf} decoder variation was proposed by \citerefs{mem_store_pscf}. 

\Gls{pscf} decoding was proposed in \cite{part_scf}, where the codeword is divided into partitions, each with its own \gls{crc}. The \gls{scf} decoding is applied to each partition separately. This technique allows to potentially exit decoding trials earlier without decoding the whole codeword. As a result, the average execution time is reduced. Also, partitioning improves the error-correction performance compared to the standard \gls{scf}.
%The technique allows to potentially exit decoding trials earlier without decoding the whole codeword, and flip single bits in each segment. The latter property improves the error-correction capability compared to \gls{scf} decoding.

The \gls{ssc} decoder was introduced in \cite{simp_sc} and uses special nodes to greatly reduce decoding latency. 
The Fast-\gls{ssc} decoder was proposed in \cite{fast_sc}. It recognizes additional types of special nodes, which further reduce decoding latency. The Fast-\gls{ssc} with these four types of special nodes was adapted to \gls{scf} decoding, and the Fast-\gls{scf} decoder was proposed in \cite{fast_sscf}. The hardware implementation of the Fast-\gls{scf} decoding was proposed in \cite{scf_eff_impl}. Then, the Fast-\gls{dscf} decoding algorithm and its hardware implementation were proposed in \cite{pract_dscf}. The Fast-\gls{scf} and Fast-\gls{dscf} decoders have shown a reduction in average execution time of \gls{scf} and \gls{dscf} decoders, with minimal impact on error-correction performance. 

\gls{scf}-based decoding and variations have variable execution time, which leads to high average execution time and latency (worst-case execution time). Therefore, there is a challenge in the realization of receivers, where fixed-time algorithms are preferred. Nevertheless, the architectural designs proposed in \cite{Giard_JETCAS_2017,scf_eff_impl,pract_dscf} showed that \gls{scf} and \gls{dscf} decoders are $4\times$ to $20\times$ more energy efficient and $2\times$ to $7\times$ more area efficient than \gls{scl} decoders with small to moderate list sizes while providing similar error-correction performance. 
%Therefore, \gls{scf}-based decoders are great candidates for 5G's \gls{mmtc} and beyond. 

The \gls{bp} decoding of polar codes~\cite{bp_intro} improves error-correction performance compared to the \gls{sc} but falls short that of the \gls{scl} decoding \cite{pol_codes_ldpc}.
The \gls{bpf} algorithm was proposed in \cite{bpf_intro} to improve the error-correction performance of the \gls{bp}. 
%Unfortunately, \gls{bpf} decoders have increased average and worst-case execution times.
%includes bits estimated as the least reliable throughout the decoding as well as according to 
In \cite{enh_bpf}, the generalized \gls{bpf} and enhanced \gls{bpf} algorithms were proposed. In these works, the set of bit-flipping candidates is constructed based on real-time estimations and according to the reliability sequence provided by the code construction.
In \cite{bpf_hw}, advanced \gls{bpf} is proposed, which utilizes a fixed critical set derived from a particular code construction algorithm. Modifications to the \gls{bpf} decoding shown in \cite{enh_bpf} and \cite{bpf_hw} achieved a reduction of the average execution time and an improvement in error-correction performance.
Nevertheless, the hardware implementations of \gls{bpf}-based decoders in \cite{enh_bpf} and \cite{bpf_hw} have shown that 
\gls{bpf}-based decoders have higher energy and area requirements compared to \gls{scl} and \gls{scf} decoder implementations when providing similar error-correction performance.

Previously, we proposed the \gls{srm} for \gls{scf}-based decoders \cite{simp_rest_mech}. \Gls{srm} conditionally restarts \gls{sc} decoding from the second half of the codeword 
if that is where the first bit-flipping candidate is located. \gls{srm} can be used with any \gls{scf}-based decoding algorithm and was shown to improve the execution-time characteristics at the cost of a small additional memory. However, that mechanism is only effective for the low-rate polar codes. The \gls{frm} was introduced in \cite{flex_rest_mech} to improve the execution-time reduction capability of the \gls{srm}. Unlike the \gls{srm}, which applies a single restart location, the \gls{frm} utilizes multiple restart locations. When applied to \gls{scf} decoder, the \gls{frm} achieves a greater average execution-time reduction compared to the \gls{srm}. However, the \gls{frm} requires significantly more additional memory than the  \gls{srm}. This will negatively impact 
area efficiency in potential hardware implementations. Moreover, the \gls{frm} selects restart locations only from the second half of the codeword, limiting its execution-time reduction capability. This could be enhanced by considering restart locations in the first half of the codeword as well.
%lead to worse area efficiency

The \gls{lrt} for \gls{scf} decoding, introduced in \cite{Giard_JETCAS_2017}, skips decoding of the parity bits that precede the first information bit. This technique was shown to reduce the average execution time of \gls{scf} decoder without requiring additional resources. However, this reduction is limited and depends solely on the number of frozen bits preceding the first information bit.

This work proposes the \gls{grm} for \gls{scf}- and Fast-\gls{scf}-based decoders for polar codes. The central idea is to skip decoding computations that are identical between the initial trial and any additional trial.
%The decoding restart is performed from the location where the course of \gls{sc} is altered for the first time, i.e., the information bit that follows the bit-flipping candidate. The computations to decode previous bits are avoided. 
%The parts of the tree in the \gls{grm} are skipped to estimate a bit-flipping candidate and all of the prior bits of each additional decoding trial. 
The decoding tree is traversed from the root along the restart path to directly estimate the restart bit. To perform a restart, the \gls{ps} bits are restored from the bit estimates stored in memory following the initial unsuccessful decoding trial. 
These \gls{ps} restorations are shown to be made by low-complexity encoding operations. The \gls{grm} does not alter error-correction performance and is adaptable to other \gls{scf}-based decoding algorithms. We show that \gls{grm} can be effectively adapted to the state-of-the-art techniques and decoders, such as \gls{scf} with \gls{lrt} \cite{Giard_JETCAS_2017}, Fast-\gls{scf} \cite{fast_sscf}  and Fast-\gls{dscf} \cite{pract_dscf}. When applied to the \gls{dscf}-3 decoder that can flip up to $3$ bits per decoding trial, our proposed \gls{grm} is shown to reduce the average execution time by $25\%$ to $60\%$.
%while requiring approximately $4\%$ additional memory compared to the original \gls{dscf}-3 decoder. 
When applied to \gls{dscf}-3 with LRT and to Fast-\gls{dscf}-3 decoders, the average execution time is reduced by $17\%$ to $23\%$ and by $15\%$ to $22\%$, respectively. For all \gls{dscf}-3-based decoders, the \gls{grm} requires approximately $4\%$ additional memory. 

The remainder of this paper is organized as follows: \autoref{sec:backgr} introduces polar codes as well as \gls{sc}, \gls{scf}- and Fast-\gls{scf}-based decoding. 
\autoref{sec:gen_rest_mech} begins with 
simulation-based statistical analysis that motivates
the use of various restart locations for decoding a codeword and then presents our proposed \gls{grm}. 
\autoref{sec:time_res_analys} first introduces the architectural execution-time model of this work, followed by the analysis of the execution-time reduction capability and additional resources required by the \gls{grm}.
In \autoref{sec:sim_res_grm}, simulation results for \gls{scf}- and Fast-\gls{scf}-based decoders with and without the \gls{grm} are provided. 
Comparisons are made in terms of error-correction performance, memory estimates, and average execution time. \autoref{sec:sim_res_grm} ends by discussing compatibility with other existing flip-based decoding algorithms. Finally, \autoref{sec:conclusion} concludes this work.

\glsreset{ps}
\section{Background}
\label{sec:backgr}
\subsection{Construction of Polar Codes}
An $(N,k)$ polar code \cite{arik_polariz} is a linear block code of length $N=2^n$, where $n\in \mathbb{N}^+$. 
The code construction is based on channel polarization, which is defined by $\bm{G}^{\otimes n}$, where $n$ is the Kronecker power of the binary kernel $\bm{G}=\left[ \begin{smallmatrix} 1 & 0 \\ 1 & 1\end{smallmatrix} \right]$. The $k$ most reliable positions are assigned to information bits, and the code rate is $R=\nicefrac{k}{N}$. 
These frozen bits, known by the decoder, are typically set to zero.
%The remaining $\left(N-k\right)$ bits, called frozen bits, are set to predefined values that are known by the decoder, typically zeros.
The encoding is performed as $\bm{x}=\bm{u} \bm{G}^{\otimes n}$, where $\bm{x}$ and $\bm{u}$ are a codeword and an input vector, respectively. The vector $\bm{u}$ contains the $k$ information bits in their predefined locations as well as frozen bits. We denote the set of frozen and information-bit indices of $\bm{u}$ by $\bm{\mathcal{A}}^C$ and $\bm{\mathcal{A}}=\{a_0, \ldots, a_j, \ldots, a_{k-1}\}$ with $a_0<\dots<a_j<\dots<a_{k-1}$. 
The bit-location reliabilities depend on the channel and conditions. In this work, we use \gls{bpsk} modulation over an \gls{awgn} channel, and the \gls{5g} construction \cite{3GPP_5G_Coding}. 

\subsection{SC Decoding}
\label{subsec:scdec}
\begin{figure}[t]
\centering
 \resizebox{1.0\columnwidth}{!}{
\input{sc_tree}}
\caption{\gls{sc} decoding tree of an $\left(8,4\right)$ polar code with a focus on a node $v$ of length $N_v=4$.}
\label{fig:pol_tree}
\end{figure}
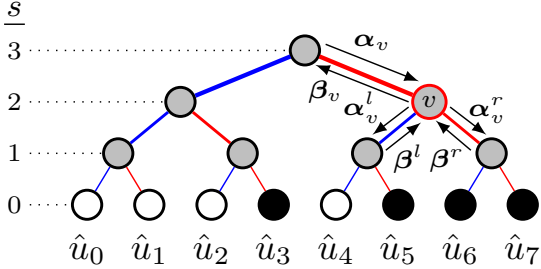
The description of \gls{sc} decoding as a binary-tree traversal was proposed in \cite{simp_sc}, where tree is traversed from top to bottom, and left to right of its branches. 
The decoding tree of an $\left(8,4\right)$ polar code with $\bm{\mathcal{A}}=\{3,5,6,7\}$ is shown  in \autoref{fig:pol_tree}, where the stages are $s\in \{n,\ldots, 0 \}$ with the root node at $s=n$. 
The received vector of channel \glspl{llr}, denoted by $\bm{\alpha}_{\text{ch}} = \{\alpha_{\text{ch}}(0), \ldots, \alpha_{\text{ch}}(N-1)\}$, is at the tree root. 
For any node, denoted by $v$, located at stage $s$, and with length $N_v=2^{s}$, the input \gls{llr} vector from the parent node is $\bm{\alpha}_{v}=\{\alpha_v\left(0\right),\ldots,\alpha_v\left(N_v-1\right)\}$, and the two input \gls{ps} vectors from its left and right child nodes are $\bm{\beta}^l=\{\beta^l\left(0\right),\ldots,\beta^l\left(\nicefrac{N_v}{2}-1\right)\}$ and $\bm{\beta}^r=\{\beta^r\left(0\right),\ldots,\beta^r\left(\nicefrac{N_v}{2}-1\right)\}$.
The left and right child nodes of $v$ receive the vectors $\bm{\alpha}_{v}^l=\{\alpha_v^l\left(0\right),\ldots,\alpha_v^l\left(\nicefrac{N_v}{2}-1\right)\}$ and $\bm{\alpha}_{v}^r=\{\alpha_v^r\left(0\right),\ldots,\alpha_v^r\left(\nicefrac{N_v}{2}-1\right)\}$, where each \gls{llr} element is calculated as follows:
\begin{align}
    \alpha_v^l(j)&=f\left(\alpha_v(j),\alpha_v\left(j+\nicefrac{N_v}{2}\right)\right),\label{eq:f}\\
    \alpha_v^r(j)&=g\left(\alpha_v(j),\alpha_v\left(j+\nicefrac{N_v}{2}\right),\beta^l(j)\right)\label{eq:g},
\end{align}
where $j\in \{0,\ldots,\nicefrac{N_v}{2}-1 \}$.
Going left, represented in blue in \autoref{fig:pol_tree}, the $f$ function is calculated as \cite{cam_schw}:
\begin{align}    f(\alpha_1,\alpha_2)=\mathrm{sgn}\left(\alpha_1 \right)  \mathrm{sgn} \left( \alpha_2 \right)  \mathrm{min}\left( \left|\alpha_1\right|,\left|\alpha_2\right| \right).\label{eq:f_function}
\end{align}
Going right, represented in red in \autoref{fig:pol_tree}, the $g$ function is calculated as:
\begin{align}
    g(\alpha_1,\alpha_2,\beta_1)=\left( 1-2\beta_1 \right) \alpha_1 + \alpha_2.
    \label{eq:g_function}
\end{align}
The bit-estimate vector $\bm{\hat{u}}=\{\hat{u}_0,\ldots,\hat{u}_{N-1}\}$ is obtained by taking a hard decision on the decision \glspl{llr}, $\bm{\alpha}_{\text{dec}} = \{\alpha_{\text{dec}}(0),\ldots, \alpha_{\text{dec}}(N-1)\}$, that reach the leaf nodes. For $i\in\bm{\mathcal{A}}$, hard decision is made based on \gls{bpsk} mapping as: 
\begin{equation}
    \label{eq:hd_llr}
    \hat{u}_i = \HD\left(\alpha_{\text{dec}}(i)\right) =
    \begin{cases}
0, ~\mathrm{if} ~ \alpha_{\text{dec}}\left(i \right) \geq 0, \\
1,  ~\mathrm{otherwise}.
\end{cases}
\end{equation}
Frozen bits are known and thus directly estimated, i.e., $\hat{u}_i=0$ if $i\in \bm{\mathcal{A}}^C$.  
In \autoref{fig:pol_tree}, the nodes represented in black and white correspond to information and frozen bits. 
The \gls{ps} vector, denoted by $\bm{\beta}_v$, is calculated from the bit estimates, and is propagated up from children to parent nodes with calculations:
%. At any node $v$, each bit of $\bm{\beta}_v$ is calculated as follows:
\begin{equation}
\label{eq:ps_combine}
\beta_v\left(j\right) = 
\begin{cases}
\beta^l\left(j\right) \oplus \beta^r\left(j\right), ~~~ \mathrm{if} ~~ j < \nicefrac{N_v}{2}, \\
\beta^r\left(j-\nicefrac{N_v}{2}\right), ~~~\;\, \,\mathrm{otherwise},
\end{cases}
\end{equation}
where $j\in \{0,\ldots,N_v-1\}$, and $\oplus$ is a bitwise XOR. This is equivalent to polar encoding.

\subsection{SCF Decoding}
\label{sec:intro_scf}
The \gls{scf} decoding algorithm is introduced in \cite{scf_intro}, where the authors observed that if the first erroneously estimated bit could be detected and corrected before resuming \gls{sc} decoding, the error-correction capability of the decoder is greatly improved. To detect a decoding failure, information bits are concatenated with an $r$-bit \gls{crc} code before applying polar encoding. The \gls{crc} bits are concatenated to the set of information bits  $\bm{\mathcal{A}}$, increasing the number of information bits of the polar code to $k_{\text{tot}}=k+r$. 
The notation $\left(N,k+r\right)$ is used in this work to indicate a polar code of length $N$, code dimension $k$, and $r$ additional \gls{crc} bits.

If a failure of the \gls{crc} is identified at the end of the initial \gls{sc} decoding trial, a bit-flipping list, denoted by $\bm{\mathcal{B}}_{\text{flip}}=\{a_0, \ldots, a_{k_{\text{tot}-1}}\}$, is constructed. The information-bit locations, that are estimated as the least reliable according to the smallest metrics $\bm{\mathcal{M}}_{\text{flip}}=\{\left|\alpha_{\text{dec}}\left(a_0\right)\right|,\ldots,\left|\alpha_{\text{dec}}\left(a_{k_{\text{tot}}-1}\right)\right|\}$ are identified. Those bit-locations are stored to $\bm{\mathcal{B}}_{\text{flip}}$ in ascending order of their corresponding metrics. 
%Therefore, $\left|\bm{\mathcal{B}}_{\text{flip}}\right|=T_{\text{max}}-1$.
When additional trials are performed, a bit-flip location  $i\in \bm{\mathcal{B}}_{\text{flip}}$ is selected. The \gls{sc} trial restarts from the beginning, and when the bit-flipping candidate is estimated, i.e., $\hat{u}_i=\mathrm{HD}\left(\alpha_{\text{dec}}\left(i\right)\right)$, the decision is inverted. The standard decoding is resumed for the remaining part of the \gls{sc} decoding in this trial. 

To constrain the latency, the maximum number of trials $T_{\text{max}}$ is defined, where $T_{\text{max}} \in \mathbb{N}^+$ and $1\leq T_{\text{max}} \leq \left( k_{\text{tot}}+1\right)$, including the initial \gls{sc} pass. Thus, the sizes of the bit-flipping list and the corresponding metrics are constrained to    $\left|\bm{\mathcal{B}}_{\text{flip}}\right|=\left|\bm{\mathcal{M}}_{\text{flip}}\right|=T_{\text{max}}-1$.
The \emph{required additional} number of trials, denoted by $\tau'$, corresponds to the trials beyond the initial decoding trial that were applied to decode a codeword. The index $t'$ denotes the current index of the additional trial throughout decoding a codeword, and $1\leq t'\leq \tau'$. Similarly, $\tau$ and $t$ denote the required number of trials and the current index of the trial, including the initial \gls{sc} pass. If the \gls{crc} still fails after running $t'=T_{\text{max}}-1$ (or $t=T_{\text{max}}$) trials,  decoding is stopped and a frame error is declared. Note that $\tau'=0$ indicates a successful decoding by the initial \gls{sc} pass. 

%Setting $T_{\text{max}}=1$ renders \gls{scf} equivalent to \gls{sc} decoding.The \emph{additional number of trials}, denoted by $\tau'$, 
%corresponds to trials beyond the initial decoding trial, 
%and $0\leq \tau' \leq T_{\text{max}}-1$. If the \gls{crc} still fails after $\tau'=T_{\text{max}}-1$ trials, the decoding is stopped and a frame error is declared. Note that $\tau'=0$ indicates a successful decoding by the initial trial. 

\subsection{Dynamic SCF Decoding}

The \gls{dscf} decoding algorithm was proposed by \cite{dyn_scf} with two major improvements to the original \gls{scf}. First, a novel metric for constructing $\bm{\mathcal{B}}_{\text{flip}}$ is derived, which increases the probability of determining the bit-flip position that results in successful decoding with smaller $\Tmax$. Second, an algorithmic improvement allowing multiple simultaneous bit flips per trial is proposed. 
%The block diagram of the \gls{dscf} decoder is shown in Figure~\ref{fig:dscf_diagr}. In this diagram, the extra steps and modifications from the \gls{scf} decoding, provided in Figure~\ref{fig:scf_diagr}, are surrounded by a dashed line. 
%\Gls{dscf} decoding algorithm is proposed in \cite{dyn_scf} with two major improvements to the original \gls{scf}. First, a more accurate metric for constructing $\bm{\mathcal{B}}_{\text{flip}}$ is derived. Second, algorithmic improvement 
%with multiple simultaneous bit flips per trial is proposed. 
The set of bit-flipping candidates is denoted by 
$\bm{\varepsilon}_{t'}=\{i_1, \ldots, i_j, \ldots,i_{\lambda}\}$, 
where $i_j\in \bm{\mathcal{A}}$ is a bit-flip location, $1 \leq j \leq \lambda$ and $\lambda$ is the current set size $\left|\bm{\varepsilon}_{t'}\right|$. Index $t'$ in $\bm{\varepsilon}_{t'}$, with $1\leq t'\leq \tau'$, denotes the current index of the additional trial.
%during the course of decoding a codeword. 
At each unsuccessful attempt $t'$, a new information index $j$ is progressively inserted to an extended set $\bm{\varepsilon}_{t'}\cup j$, increasing the size to $\lambda=\lambda+1$, and the metric $\mathcal{M}_{\text{flip}}\left(t'\right)\in \bm{\mathcal{M}}_{\text{flip}}$ is calculated as: 
\begin{equation}
\mathcal{M}_{\text{flip}}\left(t'\right) = \sum_{j\in \bm{\varepsilon}_{t'}} \left|\alpha_{\text{dec}}(j)\right| + \sum_{\substack{j \leq i_{\lambda} \\ j\in \bm{\mathcal{A}}}} \mathcal{J}\left(\alpha_{\text{dec}}\left(j\right)\right), 
\label{eq:metr_dscf_w}
\end{equation}
where $\mathcal{J}$ can be approximated according to   \cite{simp_dscf}:
\begin{equation}
\mathcal{J}\left(\alpha_{\text{dec}}\left(j\right)\right) = \begin{cases}
1.5\,, \quad \text{if} \;\; \left|\alpha_{\text{dec}}(j)\right|\leq 5.0, \\
0\,, \quad \;\;\,\text{otherwise}.
\end{cases}
\label{eq:metr_dscf_w_simp}
\end{equation}
If the metric of a newly constructed set $\bm{\varepsilon}_{t'}$ 
exceeds the largest metric in $\bm{\mathcal{B}}_{\text{flip}}$, it is discarded. If not, it is inserted to $\bm{\mathcal{B}}_{\text{flip}}$ while keeping the list sorted in ascending order according to each $\mathcal{M}_{\text{flip}}\left(t'\right)$. A set is not extended further after reaching the maximum size $\omega$. In this work, \eqref{eq:metr_dscf_w_simp} is used as it was shown to result in a negligible loss in error-correction performance \cite{simp_dscf,pract_dscf}. The decoding order $\omega$ defines the maximum size of the bit-flipping set $\bm{\varepsilon}_{t'}$ as $1 \leq \lambda \leq \omega$, i.e., $\omega=\left|\bm{\varepsilon}_{t'}\right|_{\text{max}}$ \cite{dyn_scf}. If $\omega=1$, the bit-flipping candidates are constructed at the initial unsuccessful \gls{sc} trial as in \gls{scf}, but with the metric \eqref{eq:metr_dscf_w} with $\lambda=1$, and no additional computations are performed. 

The decoder is denoted by \gls{dscf}-$\omega$ to emphasize the dependence on the parameter $\omega$. A total of $T_{\text{max}}$ trials are run with a total of $T_{\text{max}}-1$ bit-flipping sets. Note that $T_{\text{max}}$ may be greater than $(k_{\text{tot}}+1)$ when the order $\omega>1$, since in those cases, sets containing multiple bit-flips can be constructed.

\subsection{Fast Decoding and Latency-Reducing Techniques}
\label{subsec:fast_special_nodes}

As described in Section \ref{subsec:scdec}, the \gls{sc} decoding is represented as a binary-tree traversal in which each node is propagating \glspl{llr} down to the leaves, where bits are estimated. 
The construction of polar codes leads to special nodes, i.e., nodes at stage $s\geq 2$ 
with a recognizable pattern of frozen and information bits \cite{simp_sc}. A special node is denoted by $v$, with a length $N_v=2^{s_v-1}$, where $N_v\geq 4$.
 A special node does not need to be traversed until the leaf node at stage $s=0$. Instead, it can be decoded by a fast decoding technique that would significantly reduce computational complexity and decoding time.
%A special node does not need to be traversed until the leaf at $s=0$. Instead, it can be decoded by a fast decoding technique to  significantly 
%reduce execution time. 
The \gls{ssc} decoder, proposed in \cite{simp_sc}, does not traverse special nodes entirely composed of either frozen bits or information bits. These nodes are denoted by \gls{rzero} and \gls{rone} nodes. The Fast-\gls{ssc} decoder, proposed in \cite{fast_sc}, further improves the decoding time of \gls{ssc} by implementing additional types of special nodes. The most notable of these node types are the \gls{rep} and the \gls{spc} nodes. 
In this work, we also use only these four types of special nodes, since this setup is used in Fast-\gls{scf} and Fast-\gls{dscf} decoder implementations in \cite{scf_eff_impl} and \cite{pract_dscf}. The vector of \gls{ps} bits $\bm{\beta}_v$ \eqref{eq:ps_combine} is output from these nodes, and calculated according to the original schedule of \gls{sc} decoding. Next, we describe the decoding schedule of four types of special nodes used in Fast-\gls{ssc}.

\subsubsection*{R0 and R1 Nodes}
A \gls{rzero} node is entirely composed of frozen bits, and all $N_v$ bit estimates are assigned to zero. A \gls{rone} node is entirely composed of information bits, and all $N_v$ bit estimates are obtained by taking the hard decisions on the input \gls{llr} vector $\pmb{\alpha}_v$ according to \eqref{eq:hd_llr}.

\subsubsection*{\gls{rep} Node}
A \gls{rep} node of length $N_v$ is composed of a single information bit at the right-most position and $N_v-1$ frozen bits. The decision \gls{llr} $\alpha_{\text{dec}}$ for this information bit is calculated from the input vector $\pmb{\alpha}_v$ according to \cite{fast_sc}:
 \begin{equation}
    \alpha_{\text{dec}}=\sum_{j=0}^{N_v-1} \alpha_v \left(j\right).\label{eq:alpha_dec_REP}
\end{equation}
 %The information bit is estimate
%The bit estimate is obtained by taking a hard decision on $\alpha_{\text{dec}}$ \eqref{eq:alpha_dec_REP} according to \eqref{eq:hd_llr}. 
 The bit estimate is obtained by taking a hard decision on $\alpha_{\text{dec}}$ of \eqref{eq:alpha_dec_REP} according to \eqref{eq:hd_llr}. 
\subsubsection*{SPC Node}
An \gls{spc} node of length $N_v$ is composed of a single frozen bit at the left-most location and  $N_v-1$ information bits. Decoding is made by taking hard decisions on all elements of $\pmb{\alpha}_v$ and satisfying parity constraint $p=0$.
The bit indices in the \gls{spc} node are denoted by $i\in\{0,\ldots, N_v-1\}$, where the frozen bit is at $i=0$.
By denoting the bit indices in \gls{spc} node by $i\in\{0,\ldots, N_v-1\}$,  where the frozen bit is at $i=0$.
The \gls{spc} node decoding can be defined according to \cite{fast_sc} and \cite{scf_eff_impl} as follows:
%\newline
\begin{equation}
    %\label{eq:spc_dec}
    \hat{u}_i = 
    \begin{cases}
0, ~\mathrm{if} ~ i=0, \\
0, ~\mathrm{if} ~ i>0 ~ \mathrm{and} ~ \alpha_{v}\left(i \right) \geq 0, \\
1, ~  \mathrm{if} ~ i>0 ~ \mathrm{and} ~ \alpha_{v}\left(i \right) < 0.
\end{cases}
\label{eq:spc_hd}
\end{equation}
The hard decisions in \eqref{eq:spc_hd} must satisfy the parity constraint as follows:
\begin{equation}
\label{eq:spc_par}
  p = \bigoplus_{i=0}^{N_v-1} \hat{u}_i.
\end{equation}
If parity constraint $p=1$ in \eqref{eq:spc_par}, 
the least reliable information bit according to $\left|\alpha_{v}\left(i\right)\right|$ is flipped. 

The Fast-\gls{ssc} trial is denoted by \fssc.
The decoding tree of the $\left(8,4\right)$ polar code shown in Figure~\ref{fig:pol_tree} can now be simplified. The resulting {\fssc} decoding tree composed of one \gls{rep} node and one \gls{spc} node of lengths $N_v=4$ is shown in \autoref{fig:fastssc-tree}.

\subsubsection*{Latency-Reducing Technique}

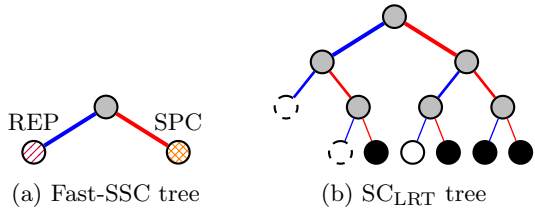
\begin{figure}[t]
    \centering
     \subfloat[Fast-\gls{ssc} tree]{\label{fig:fastssc-tree}\rotatebox{90}{\input{fastssc_tree_sscflip}}}
    \hspace{20pt}
    \subfloat[$\scLRT$ tree]{\label{fig:sc_lrt_tree}\rotatebox{90}{\input{sc_tree_sscflip}}}   
  \caption{Fast-\gls{ssc} \cite{fast_sc}  and  $\scLRT$ \cite{Giard_JETCAS_2017} decoding tree representations of an $(8,\,4)$ polar code.}
  \label{fig:fast_ssc_lrt}
\end{figure}

In their study, \citep{Giard_JETCAS_2017} proposed an implementation of the \gls{scf} decoder, where each \gls{sc} trial is modified by the \gls{lrt}. This technique is entirely based on skipping traversal of the left-most frozen bits in the decoding tree, meaning that it resumes the decoding at the first information-bit location $a_0$.
The \gls{sc} trial with the \gls{lrt} is denoted by $\scLRT$. 
The $\scLRT$ decoding tree is depicted in 
 \autoref{fig:sc_lrt_tree}, where the nodes represented in black and white correspond to information and frozen bits. The nodes corresponding to the frozen bits at the \gls{lhs} of the first information bit are visualized by the dashed line. These nodes are not traversed in $\scLRT$. 

%In their study, \citep{Giard_JETCAS_2017} proposed an implementation of the \gls{scf} decoder, where each \gls{sc} trial is modified by the \gls{lrt}. 
%This technique is entirely based on skipping traversal of the left-most frozen bits in the decoding tree, meaning that it resumes the decoding at the first information-bit location $a_0$.
%The \gls{sc} trial with the \gls{lrt} is denoted by $\scLRT$. 
%This technique resumes decoding from the first information bit-location $a_0$. 
%The \gls{sc} trial with the \gls{lrt} is denoted by $\scLRT$. The $\scLRT$ decoding tree is depicted in \autoref{fig:sc_lrt_tree}.

%Contrary to \gls{sc} decoding, {\fssc} and $\scLRT$ have variable execution times when pattern of frozen and information bits changes. 
In this work, \gls{sc}, $\scLRT$ \cite{Giard_JETCAS_2017} and {\fssc} \cite{fast_sc} serve as the baseline algorithms 
``dec'' for \gls{scf}-based decoders. The proposed \gls{grm}, which simplifies additional decoding trials, can be implemented over a modified \gls{sc} baseline algorithm. %We use the notation $\text{dec}\in\{\text{SC},\scLRT,\text{\fssc}\}$.

\section{Generalized Restart Mechanism}
\label{sec:gen_rest_mech}
This section begins with a statistical analysis of the bit-flipping candidates in a codeword of \gls{scf}-based decoding.
Then, the \gls{grm}, a novel execution-time reduction for flip decoders with any \gls{sc} baseline algorithm, is described with an illustrative example as support. 

\subsection{Distribution of the Bit-Flipping Candidates}
\label{sec:statist_analys}
A statistical analysis is carried out to analyze the \gls{pmf} of the bit-flipping candidates. This is beneficial to understand the limitations of the previously proposed \gls{srm} \cite{simp_rest_mech}, and to motivate the use of the proposed \gls{grm}.
In other existing works, the analyses are mainly focused on identifying unsuccessfully decoded codewords by \gls{scf} \cite{eis_scf,thr_scf} and \gls{dscf} \cite{earl_stop_dscf} decoders, throughout the simulations. By identifying these codewords, decoding is terminated at earlier decoding trials. However, this approach negatively impacts the error-correction performance. In this work, we focus on both successfully and unsuccessfully decoded codewords as to improve execution-time characteristics without affecting original error-correction performance. 

\begin{figure*}[t]
 \centering	 
 \resizebox{1.0\textwidth}{!}{\input{Flip_Distrib_dscf3_r025.tex}}
  \caption{\Gls{pmf} of the information-bit location $a_j\in \bm{\mathcal{A}}$ being the first bit-flipping candidate $i_1= \varepsilon_{t'}\left(0\right)$ under \gls{dscf}-3 decoding for the code rate $R=\nicefrac{1}{4}$. $\mathbb{P}_{\text{LHS}}$ and $\mathbb{P}_{\text{RHS}}$ are inside each plot.} 
 \label{fig:pdf_flip_dscf3_N1024_r025}
\end{figure*}
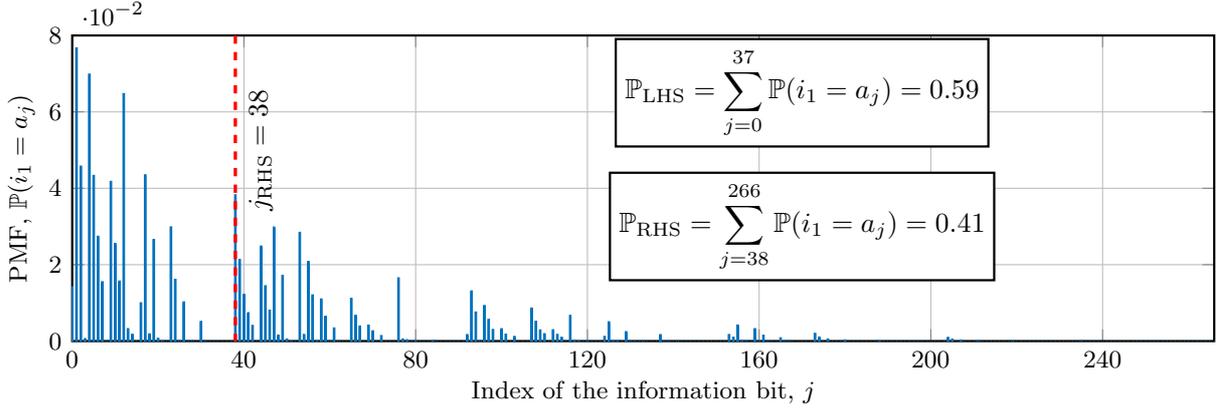

For this analysis, \gls{5g} polar codes~\cite{3GPP_5G_Coding} of length $N=1024$  with rate $R = \nicefrac{1}{4}$ and a \gls{crc} of $r=11$ bits with  polynomial $z^{11}+z^{10}+z^{9}+z^{5} + 1$ are used. 
%Simulations are made over the \gls{awgn} channel with \gls{bpsk} modulation.
The \gls{bpsk} modulation is used over an \gls{awgn} channel. 
Simulations are performed by estimating a minimum of $C=2\cdot 10^5$ codewords or until $2000$ frame errors are observed for the target \gls{fer} of $10^{-2}$.
%Simulations are for a minimum of $C=2\cdot 10^5$ codewords, and are run until at least $2000$ errors are observed for the target \gls{fer} of $10^{-2}$.
The multi-bit flip \gls{dscf}-$\omega$ decoder  with $\omega=3$ and $T_{\text{max}}=301$ is simulated.
The value of $T_{\text{max}}$ is selected to achieve an error-correction performance that is close to the genie-aided decoder \cite{dyn_scf} at the \gls{fer} of $10^{-2}$.
The \gls{pmf} distribution includes only the first bit in a bit-flipping set, i.e, bit $i_1=\varepsilon_{t'}\left(0\right)$, for the set of information bits $\bm{\mathcal{A}}=\{a_0, \ldots, a_j, \ldots, a_{k_{\text{tot}}-1}\}$, such that $0\leq j\leq k_{\text{tot}}-1$. The probability that $a_j$ is the first bit to be flipped during an additional trial is further denoted by $\mathbb{P}\left(i_1=a_j\right)$. 

In our previously proposed \gls{srm} \cite{simp_rest_mech}, additional trials are restarted on the \gls{lhs}, i.e., at $0$ or on the \gls{rhs}, i.e., at $\nicefrac{N}{2}$ of the decoding tree depending on the location of $i_1$. The probability of a restart from the \gls{lhs} in the \gls{srm} \cite{simp_rest_mech} is 
\begin{align}  \mathbb{P}_{\text{LHS}}=\mathbb{P}\left(i_1<\frac{N}{2}\right)=\sum_{j=0}^{j_{\text{RHS}}-1} \mathbb{P}(i_1=a_j),
\end{align}
where $j_{\text{RHS}}$ is location of the first information bit on the \gls{rhs} of the decoding tree.
Likewise, the probability to restart from the \gls{rhs} in the \gls{srm} is defined as: 
\begin{equation}    
\begin{aligned}
\mathbb{P}_{\text{RHS}}&=\mathbb{P}\left(i_1\geq\frac{N}{2}\right) = \\
&=\sum_{j=j_{\text{RHS}}}^{k_{\text{tot}}-1} \mathbb{P}(i_1=a_j)=1-\mathbb{P}_{\text{LHS}}.
\end{aligned}
\end{equation}

\begin{table}[t]
    \centering
    \caption{Summary of the \gls{pmf} characterizing \gls{lhs} and \gls{rhs} of the decoding tree for different code rates.}
    \setlength{\tabcolsep}{4mm} 
    \renewcommand{\arraystretch}{1.3} 
    \begin{tabular}{cccc}
       \toprule
       Code rate& $\mathbb{P}_{\text{LHS}} \left(\%\right)$  & $\mathbb{P}_{\text{RHS}} \left(\%\right)$ & $j_{\text{RHS}}$\\
      % & $\%$ & $\%$ & \\
       \midrule
       $R=\nicefrac{1}{8}$ & $29$  & $71$ & $10$  \\ 
       $R=\nicefrac{1}{4}$ & $59$  & $41$ & $38$ \\
       $R=\nicefrac{1}{2}$ &  $90$ & $10$ & $144$ \\       
       \bottomrule         
    \end{tabular}    
    \label{tab:pmf_dif_rate}
\end{table}

The \gls{srm} permits to improve the average execution time when restart from the \gls{rhs} has a high probability. \autoref{fig:pdf_flip_dscf3_N1024_r025} 
depicts the resulting \gls{pmf} for rate $R=\nicefrac{1}{4}$.
The red dashed line indicates the first information index at the \gls{rhs} tree, i.e., $j_{\text{RHS}}$, and the probabilities $\mathbb{P}_{\text{LHS}}$ and  $\mathbb{P}_{\text{RHS}}$ are overlayed. \autoref{tab:pmf_dif_rate} summarizes $\mathbb{P}_{\text{LHS}}$, $\mathbb{P}_{\text{RHS}}$ and $j_{\text{RHS}}$ for rates $R\in\{\nicefrac{1}{8},\nicefrac{1}{4},\nicefrac{1}{2}\}$.
Looking at results, the correlation between 
$j_{\text{RHS}}$ and $\mathbb{P}_{\text{LHS}}$ (or $\mathbb{P}_{\text{RHS}}$) is evident. Notably, for the rate $R=\nicefrac{1}{8}$, the tree-division index is very low such that $j_{\text{RHS}}=10\ll k_{\text{tot}}$. That pushes the bit-flipping candidates more towards the \gls{rhs}, resulting in a smaller $\mathbb{P}_{\text{LHS}}$. As the code rate increases, the bit-flipping candidates tend to be located on the \gls{lhs}, resulting in a greater $\mathbb{P}_{\text{LHS}}$.
This is why the \gls{srm} was shown to be more effective for a low code rate such as $R=\nicefrac{1}{8}$ compared to a high code rate such as $R=\nicefrac{1}{2}$ \cite{simp_rest_mech}. 
A mechanism that would be able to restart anywhere  in a codeword would further reduce the average execution time of \gls{scf}-based decoding and greatly improve the \gls{srm} \cite{simp_rest_mech}.
\glsreset{grm}
Hence, the \gls{grm} is proposed with this motivation and is described in the following section.

\subsection{The Generalized Restart Mechanism with SC as Baseline}
\label{sec:grm_descr}
At additional decoding trials, the \gls{scf}-based decoder starts by traversing the whole decoding tree to estimate each information bit until the one that needs to be flipped, at index $i_1= \varepsilon_{t'}\left(0\right)$. After flipping this bit, the standard course of decoding is resumed for the remaining part of the decoding tree if $\left|\bm{\varepsilon}_{t'}\right|=1$.
If $\left|\bm{\varepsilon}_{t'}\right|>1$, at least one additional bit flip will be performed in the remaining part of the decoding tree. The \gls{grm} aims to reduce the number of computations for each additional decoding trial $t'$. This results in a reduced decoding time of $t'$, and ultimately, a reduced average execution time for decoding. 
%The \gls{grm} aims at reducing the number of computations of \gls{scf}-based decoder which will lead to a reduced decoding time.

%First, a general definition of the proposed \gls{grm} is given.
First, a fundamental definition of the proposed \gls{grm} is provided in Definition~\ref{def:grm}.
Second, a definition of the restart path of \scmodif is provided in Definition~\ref{def:rest_path}. 
 This is followed by a definition of the partial-sum restoration provided in Definition~\ref{def:ps_rest}. %Then, a definition of the partial-sum restoration is provided in Definition~\ref{def:ps_rest}. 

\begin{definition}[The \gls{grm} of SCF-based decoder]\label{def:grm}
\normalfont{
For an additional trial $t'$, the flip decoding algorithm with the proposed \gls{grm} skips the non-negligible part of the decoding tree by restarting at location $\psi_{t'}$.
 This modified \gls{sc} trial is denoted by  
 \scmodif, indicating the restart location $\psi_{t'}$ and the bit-flipping set $\pmb{\varepsilon}_{t'}$.
 The savings in terms of computations and decoding time result from storing the $N$ bit estimates $\bm{\hat{u}}$ of the initial \gls{sc} trial. These bit estimates are stored to $\bm{\hat{u}_{\text{rest}}}$.
 } 
\end{definition}
 When performing an additional trial of the \gls{sc} decoding with the set of bit-flipping candidates $\bm{\varepsilon}_{t'}$, the additional trial does not differ from the initial \gls{sc} pass until the first flipping location $i_1=\varepsilon_{t'}\left(0\right)$.
Hence, all bit estimates $\{\hat{u}_0,\ldots,\hat{u}_{i_1-1}\}$  are identical to the estimations of the initial \gls{sc} trial. Their storage in $\bm{\hat{u}_{\text{rest}}}$ permits to assign 
\begin{align}
    \hat{u}_j=\hat{u}_{\text{rest}}(j) 
    \label{eq:prior_i1},
\end{align} 
where $0\leq j\leq i_1-1$. For position $i_1$, the value of the bit $\hat{u}_{i_1}$ is flipped with respect to $\hat{u}_{\text{rest}}(i_1)$.
Hence, no computations are required to determine the value of the bit $\hat{u}_{i_1}$.
For positions greater than $i_1$, the bit estimates are known if they are frozen and unknown if the bit belongs to $\bm{\mathcal{A}}$.

Hence, no computations are required until the first information bit %that follows $i_1$ 
on the \gls{rhs} of $i_1$ is met.
The tree traversal to compute $\alpha_{\text{dec}}\left(\psi_{t'}\right)$ to estimate $\hat{u}_{\psi_{t'}}$ corresponds to the \emph{restart path} of \scmodif.
After the restart path, the \gls{sc} trial with $\bm{\varepsilon}_{t'}$ is resumed for the remaining part of the decoding tree to estimate bits $\left\{\hat{u}_{\psi_{t'}+1}, \ldots,\hat{u}_{N-1}\right\}$. 

%\subsection{The Restart Path in an Additional Trial}
%\label{sec:grm_rest_path}
%The restart path of \scmodif is defined as follows:
\begin{definition}[Restart path of \scmodif]\label{def:rest_path}
    \normalfont{
    In an additional trial $t'$, the restart path is the tree traversal from the root $s=n$ to the leaf $s=0$, leading to the decision \gls{llr} $\alpha_{\text{dec}}(\psi_{t'})$. It allows to compute $\hat{u}_{\psi_{t'}}=\HD\left(\alpha_{\text{dec}}\left(\psi_{t'}\right)\right)$ \eqref{eq:hd_llr}. The type of function \eqref{eq:f_function}--\eqref{eq:g_function}, performed at each decoding stage $\varsigma \in \{n-1,\ldots, 0\}$ is determined by the vector $\bm{\mathcal{H}}_{\psi_{t'}}$, which is the binary representation of $\psi_{t'}$.}
\end{definition} 
The binary representation of $\psi_{t'}$ is denoted by $\bm{\mathcal{H}}_{\psi_{t'}}=\left\{\mathcal{H}_{\psi_{t'}}\left(0\right), \ldots, \mathcal{H}_{\psi_{t'}}\left(n-1\right)\right\}$, where the \gls{lsb} is on the right. 
At the decoding stage $\varsigma\in\{n-1,\ldots,0\}$ of the restart path, the bit $\mathcal{H}_{\psi_{t'}}\left(n-1-\varsigma\right)$ is used as flag to determine whether an $f$-function or a $g$-function is performed to return the \gls{llr} vector $\pmb{\alpha}_\varsigma=\left\{\alpha_\varsigma(0),\dots,\alpha_\varsigma(2^\varsigma-1)\right\}$ that will be used at stage $\varsigma-1$. 
If $\mathcal{H}_{\psi_{t'}}\left(n-1-\varsigma\right)=0$,  $f$-function is performed at decoding stage $\varsigma$ and 
\begin{align}
    \alpha_\varsigma(j) = f\left(\alpha_{\varsigma+1}(j),\alpha_{\varsigma+1}(j+2^\varsigma)\right),
\end{align}
where $j\in\{0, \dots, 2^\varsigma-1\}$ \eqref{eq:f}.
If $\mathcal{H}_{\psi_{t'}}\left(n-1-\varsigma\right)=1$,  $g$-function is performed at decoding stage $\varsigma$.
%If a $g$-function is required at stage $\varsigma$ at the restart path, 
In this case, a segment of $2^{\varsigma}$ partial-sums, denoted by $\bm{\beta}_{\varsigma}$, is required and is obtained by the partial-sum restoration. The partial-sum restoration is explained in Definition~\ref{def:ps_rest}.

The resulting vector $\bm{\alpha}_\varsigma$ is retrieved as
\begin{align}
    \alpha_\varsigma(j) = g\left(\alpha_{\varsigma+1}(j),\alpha_{\varsigma+1}(j+2^\varsigma),\beta_\varsigma(j)\right),
\end{align}
where $j\in\{0, \dots, 2^\varsigma-1\}$ \eqref{eq:g}.
Note that the computation of vectors $\bm{\alpha}_\varsigma$ do not need extra memory and are being stored in the memory for \gls{sc} intermediate \glspl{llr} $\bm{\alpha}_{\text{int}}$. 
%, composed of $N-1$ \glspl{llr}.}

\begin{table*}[t]
    \centering    
  \small
\parbox{0.87\textwidth}{
\caption{Summary of the key differences between execution-time reduction mechanisms for \gls{scf}-based decoding.}
 \label{tab:sum_dif_schemes}}
    \setlength{\tabcolsep}{3pt} % Default value: 6pt
     \renewcommand{\arraystretch}{1.1} % 
   %\resizebox{0.97\textwidth}{!}{
\begin{tabular}{ccccccc}
        \toprule
         \multirow{2}{*}{Reference} & Baseline & Additional trial & Restart & Restart index  & Memory & \multirow{2}{*}{Label} \\
          & algorithm dec & decoder & Mechanism & (add. trial) & overhead & \\
          %\hline\hline
          \midrule
         \cite{scf_intro,dyn_scf} & \gls{sc} & $\text{SC}\left(\bm{\varepsilon}_{t'} \right)$ & \textbf{--} & $0$ & $0\%$ & SCF, DSCF-$\omega$\\
         \midrule
         \cite{Giard_JETCAS_2017} & $\scLRT$ & $\text{SC}\left(a_0, \bm{\varepsilon}_{t'} \right)$ & \textbf{--} & $a_0$ & $0\%$ & \gls{lrt}\\
         \midrule
         \cite{fast_sscf,pract_dscf} & Fast-SSC & FSSC$(\bm{\varepsilon}_{t'})$ & \textbf{--} & $a_0$ & $0\%$ & FAST\\
         \midrule
         \cite{simp_rest_mech} & SC & SC(0 or $\nicefrac{N}{2},\bm{\varepsilon}_{t'})$ & \gls{srm} & 0 or $\nicefrac{N}{2}$ & $4\%$ to $7\%$ & SRM\\
         \midrule
         \multirow{3}{*}{This work} & SC &  \scmodif & \gls{grm} & $\psi_{t'}$ & $4\%$ to $7\%$ & \gls{grm}\\
          &  $\scLRT$ & \scmodif &  \gls{grm} & $\psi_{t'}$ & $4\%$ to $7\%$ & \gls{lrt}+\gls{grm}\\
          &  Fast-SSC & \fsscmodif & \gls{grm} & $\psi_{t'}$ & $4\%$ to $7\%$ & FAST+GRM\\
          \bottomrule
    \end{tabular}  % }
   
\end{table*}

\begin{definition}[Partial-sum restoration]\label{def:ps_rest}
    \normalfont{
    At any decoding stage $\varsigma\in\{n-1,\ldots,0\}$, where $g$-function is to be performed, the \acrfull{ps} restoration corresponds to the computations of $2^{\varsigma}$ \gls{ps} bits $\pmb{\beta}_{\varsigma}$ used in the $g$-function on the basis of a segment of $2^{\varsigma}$ bit estimates $\pmb{\eta}_\varsigma$. The start index of the segment $\pmb{\eta}_{\varsigma}$ is obtained by subtracting an offset from $\psi_{t'}$. This  offset is further denoted by $\phi_{\varsigma}$ and can be obtained as follows:}
    \begin{equation}
        \phi_{\varsigma} = \sum_{s=0}^{\varsigma} \left(\mathcal{H}_{\psi_{t'}}\left(n-1-s\right) \cdot 2^s\right).
        \label{eq:offs_bit_est}
    \end{equation}
    \normalfont{A polar encoder of size $2^\varsigma$ \eqref{eq:ps_combine} on  $\bm{\eta}_\varsigma$ allows to retrieve $\bm{\beta}_\varsigma$.}
\end{definition}
The partial-sum restoration in the architectural model is discussed in more detail in Section \ref{sec:add_req_grm_ps_restore}.

\begin{figure*}[t]
%\centering	
\centering\resizebox{0.7\textwidth}{!}
{\input{Saved_tree_branches}}
\caption{The modified \gls{sc} trial  \scmodif with the  bit-flipping index $i_1=9$ and the restart location $\psi_{t'}=11$, in \gls{scf} with the \gls{grm} for a $\left(16,8\right)$ polar code. The elements in red are computed during the restart path of \scmodif.}
\label{fig:illustr_grm_rest_N16}
\end{figure*}
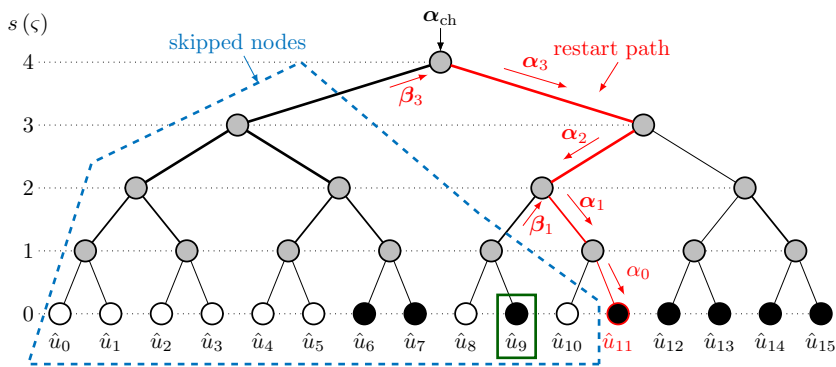

The modified \gls{sc} trial in flip-based decoder with the \gls{grm}, \scmodif, is illustrated in \autoref{fig:illustr_grm_rest_N16} with a $\left(16,8\right)$ code defined by $\bm{\mathcal{A}}=\{6,7,9,11,12,13,14,15\}$. In this example, the bit-flipping location is $i_1=a_2=9$, and the restart location is set to the next information bit with index $\psi_{t'}=a_3=11$.
Bit estimates at the \gls{lhs} of $\hat{u}_{9}$, i.e, bits $\{\hat{u}_0,
 \ldots, \hat{u}_{8}\}$, are restored from the  memory $\bm{\hat{u}}_{\text{rest}}$ according to \eqref{eq:prior_i1}.
 Then, the bit $\hat{u}_{9}=\hat{u}_{i_1}$ is restored from $\hat{u}_{\text{rest}}\left(9\right)$ and flipped, i.e., $\hat{u}_{9}=\hat{u}_{\text{rest}}(9)\oplus 1$.
 The bit $\hat{u}_{10}$ is assigned to $0$ since $10\in\bm{\mathcal{A}}^C$.
As a result of applying the \gls{grm}, the parts of the tree  to estimate the bits $\{\hat{u}_0,\ldots,\hat{u}_{\psi_{t'}-1}\}$, surrounded by the dashed blue line in  \autoref{fig:illustr_grm_rest_N16}, are skipped.
The restart path of \scmodif, which connects the root of the tree at stage $s=n=4$ with the leaf $\hat{u}_{11}$ at stage $s=0$, is highlighted in red in \autoref{fig:illustr_grm_rest_N16}.
During the traversal, at the decoding stages $\varsigma\in \{n-1,\ldots,0\}$ and given $\bm{\mathcal{H}}_{\psi_{t'}}=\{1,0,1,1\}$, $g$-functions are performed at stages $\varsigma\in\{3,1,0\}$, and  $f$-functions at stage $\varsigma=2$. 
%The standard course of \gls{sc} decoding is resumed for the remaining part of the decoding tree upon traversing the restart path and estimating $\hat{u}_{11}$. 

In more detail, at the root, the decoding stage is $\varsigma=3$,  $g$-functions are performed with $\bm{\alpha}_4\triangleq\bm{\alpha}_{\text{ch}}$ and $\bm{\beta}_3$ (yet to be restored).
The \gls{ps} values $\bm{\beta}_3$ are restored on the basis of $2^3=8$ bit estimates. The offset subtracted from $\psi_{t'}$ is $\phi_3=8+2+1=11=\psi_{t'}$ \eqref{eq:offs_bit_est} leading to the segment $\bm{\eta}_3=\{\hat{u}_0,\dots,\hat{u}_7\}$.
The $g$-functions return a vector of \glspl{llr} $\bm{\alpha}_3$, which will be used for the following $f$-functions at $\varsigma=2$.
No \gls{ps} values are required \eqref{eq:f_function}, and  $\bm{\alpha}_2$ is returned. 
Next, $g$-functions are performed at the decoding stage $\varsigma=1$ to compute $\bm{\alpha}_1$.
The segment $\bm{\beta}_1$ is restored on the basis of $\bm{\eta}_1=\{\hat{u}_8,\hat{u}_9\}$, since $\phi_1=2+1=3$ \eqref{eq:offs_bit_est}.
The vector $\bm{\alpha}_1$ is returned and used to perform the following $g$-function.
Finally, $g$-function is performed at the stage $\varsigma=0$. 
For the \gls{ps}, the segments verify $\bm{\eta}_0=\bm{\beta}_0=\hat{u}_{10}$, since the offset is $\phi_0=1$ \eqref{eq:offs_bit_est}.
The $g$-function corresponds to 
\begin{align}
    \alpha_0=\alpha_{\text{dec}}\left(11\right) = g\left(\alpha_1(0),\alpha_1(1),\hat{u}_{10}\right),
\end{align} 
which allows to estimate $\hat{u}_{11}$ and conclude the restart path. The standard course of \gls{sc} decoding is then resumed for the remaining part of the decoding tree.

\subsection{GRM and Other Latency-Reduction Techniques}
\label{sec:incorp_grm}
The proposed \gls{grm} is flexible to apply with other existing techniques that reduce the latency of the \gls{sc} decoding. As a result of this combination, the average execution time of \gls{scf}-based decoders can be further reduced.
Next, two modified versions of \gls{sc}, namely $\scLRT$ and Fast-\gls{ssc}, are considered.

\subsubsection*{Latency-Reducing Technique}
 
In \gls{scf} with \gls{lrt}, the restart location $a_0$ is applied for both initial and additional trials.
For any additional trial $t'$, the decoding can be derived by $\scLRT=\text{SC}\left(a_0, \bm{\varepsilon}_{t'} \right)$, i.e., $a_0$ is the restart location, and $\bm{\varepsilon}_{t'}$ is the set of bit-flipping candidates.  The notation $\text{SC}\left(a_0, \bm{\varepsilon}_{t'} \right)$ follows the same logic that for the modified \gls{sc} trial \scmodif implied by the proposed \gls{grm}.
Our proposed \gls{grm} is compatible with the \gls{lrt}. 
Namely, embedding the \gls{grm} with the \gls{lrt} comes down to performing the modified \gls{sc} trial \scmodif for the additional trials rather than $\text{SC}\left(a_0, \bm{\varepsilon}_{t'} \right)$, reducing the decoding time for additional trials. 

\subsubsection*{Fast-\gls{scf} Decoding}

In a Fast-\gls{scf} decoder with Fast-\gls{ssc} as the baseline algorithm, the additional trial $t'$ is denoted by \fsscbitflip.
%The {\fssc} for any additional trial $t'$ in Fast-\gls{scf} with the bit-flipping set $\bm{\varepsilon}_{t'}$ is denoted by \fsscbitflip.
This technique resumes decoding from the first information bit-location $a_0$ in the codeword.  
Similarly to the standard \gls{sc}, the bit estimates prior to $i_1$ are identical to the initial trial. Therefore, the \gls{grm} can also be applied to Fast-\gls{scf} decoding. At additional trial $t'$, the modified Fast-\gls{ssc} trial is denoted by \fsscmodif. The restart path of \fsscmodif is made with the following assumptions:
\begin{enumerate}
    \item If the bit-flipping index $i_1=\varepsilon_{t'}(0)$ belongs to a special node, \fsscmodif begins with this special node.     
    \item If $i_1=\varepsilon_{t'}(0)$ is located outside of a special node, the \fsscmodif begins by decoding the first information bit on the \gls{rhs} of $i_1$, same as for $\text{SC}\left(\psi_{t'}, \bm{\varepsilon}_{t'} \right)$.
    \begin{itemize}
        \item If $\hat{u}_{\psi_{t'}}$ falls in a special node, the \fsscmodif begins with this special node.
    \end{itemize}
\end{enumerate}

The same assumptions are applied to Fast-\gls{dscf} decoding.
The Fast-\gls{scf} and Fast-\gls{dscf} with the \gls{grm} will further reduce the average execution time of the standard \gls{scf} and \gls{dscf} decoders.
\autoref{tab:sum_dif_schemes} summarizes execution-time reduction mechanisms including the \gls{grm} with different baseline algorithms.

\glsreset{ps}
\section{Time and Resource Analyses of GRM}
\label{sec:time_res_analys}
\begin{figure*}[t]
 \centering	
 \begin{subfigure}[t]{1.0\columnwidth}
 \centering\resizebox{1.0\columnwidth}{!} {\input{Funct_save_GRM_P}}\vspace{-0.15cm}
    \caption{\gls{grm} with various $P$ values.}
     \label{fig:tim_save_grm_p}
 \end{subfigure}
 \begin{subfigure}[t]{1.0\columnwidth}
 \centering\resizebox{1.0\columnwidth}{!}  {\input{Funct_save_GRM_SRM}}\vspace{-0.15cm}
     \caption{\gls{grm} Vs. \gls{srm} \cite{simp_rest_mech} and \gls{lrt}  \cite{Giard_JETCAS_2017} with $P=64$.}
     \label{fig:tim_save_grm_srm}
 \end{subfigure}
  \parbox{0.95\textwidth}{\caption{Execution-time reduction of \scmodif estimated by \eqref{eq:tot_saved_cc} for a $\left(1024,512+11\right)$  code with $\psi_{t'}=a_j\in\bm{\mathcal{A}}$.
  Latency of the \gls{sc} decoder included for reference.} 
 \label{fig:time_save_grm_all}}
\end{figure*}
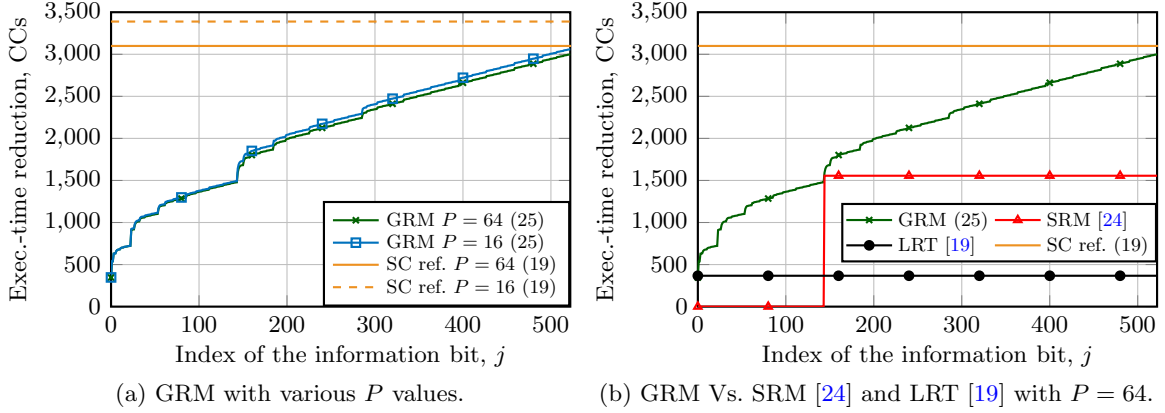

This section begins with an introduction to the architectural time model. Based on this model, the execution-time reduction capability of {\scmodif} in the \gls{grm} is derived, and a comparison is made with other existing time-reduction techniques, where various baseline algorithms are used. 
Then, a model describing the memory resources of \gls{scf} with the \gls{grm}, and the resources of partial-sum restoration are provided. 

This section begins with the introduction to the architectural time model. Then, the execution-time reduction capability of \scmodif in the \gls{grm} is derived, and comparison is made with other existing time-reduction techniques, where various baseline algorithms are used.  
Finally, a model describing the memory resources of \gls{scf} with the \gls{grm}, and the resources of partial-sum restoration are provided. 

\subsection{Architectural Execution-Time Model}
The architectural execution-time model used in this work is based on the semi-parallel \gls{sc} decoder implementation from \cite{semi_par_sc}. 
%This model is used throughout the doctoral study and in this chapter.
The execution-time model is derived for three baseline decoding algorithms: \gls{sc}, $\scLRT$, and \fssc. 
For each node in the decoding tree (\autoref{fig:pol_tree}), the \gls{llr} values are calculated in one \gls{cc} with a limited number of processing elements, denoted by $P$.
The \gls{ps} bits are calculated according to \eqref{eq:ps_combine} with $P$ parallel XOR blocks that generate $2P$ bits in one \gls{cc} \cite{fast_sc}. 
Based on this model, the execution time of the standard \gls{sc} trial $\mathcal{L}_{\text{SC}}$, expressed in \glspl{cc}, is calculated as:
% The architectural execution-time model is studied for three baseline algorithms, \gls{sc}, $\scLRT$, and \fssc, which is based on the semi-parallel \gls{sc} decoder implementation \cite{semi_par_sc}.
% For each node in the decoding tree (\autoref{fig:pol_tree}), \gls{llr} values are calculated in one \gls{cc} with a limited number of processing elements, denoted by $P$.
% The \gls{ps} bits are calculated according to \eqref{eq:ps_combine} with $P$ parallel XOR blocks that generate $2P$ bits in one \gls{cc} \cite{fast_sc}. 
% The execution time of a standard \gls{sc} decoding trial is denoted by $\mathcal{L}_{\text{SC}}$, expressed in \glspl{cc}, and summarized as:
\begin{equation}
\mathcal{L}_{\text{SC}} = \mathcal{L}_{\alpha}(N)+\mathcal{L}_{\beta}(N),
\label{eq:lat_sc}
\end{equation}
where the components $\mathcal{L}_{\alpha}(N)$  and $\mathcal{L}_{\beta}(N)$ denote execution times of \gls{llr} and \gls{ps} calculations for the polar code of length $N$. Each component is calculated by the following equations:
\begin{align}   
\mathcal{L}_{\alpha}(N) &=  2N + \frac{N}{P}\cdot \log_2{\left(\frac{N}{4P}\right)}, 
\label{eq:lat_sc_f_g} \\
\mathcal{L}_{\beta}(N) &=  \sum_{s=1}^{n-1}  \left( 2^{n-s}-1 \right) \cdot \left \lceil \frac{2^s}{2P} \right \rceil ,
\label{eq:lat_sc_ps}
\end{align}
where $\mathcal{L}_{\alpha}(N)$ \eqref{eq:lat_sc_f_g} was given in \cite{semi_par_sc}, while $\mathcal{L}_{\beta}(N)$ \eqref{eq:lat_sc_ps} is derived by using the results of  \cite{fast_sc}. 
The term $\left( 2^{n-s}-1 \right)$ in \eqref{eq:lat_sc_ps} indicates the number of nodes at tree stage $s$, excluding the right-most node.
The term $\nicefrac{2^s}{2P}$  corresponds to the number of \glspl{cc} required to compute \gls{ps} for each node. For the values of $P\geq \nicefrac{N}{4}$, \eqref{eq:lat_sc_ps} is simplified to $\mathcal{L}_{\beta}(N)=N-n-1$, as derived by us in \cite{simp_rest_mech}.
Finally, note that the ``-1'' in \eqref{eq:lat_sc_ps} means that the final computations (going up on the right edges of the last leaf in the tree) are not performed. 

%We further denote by  
%the execution time of \gls{ps} that include the final computations, which is calculated as:
The execution time of \gls{ps} bits, which include the final computations, is denoted by $\mathcal{L}_{\beta_{\text{full}}}(N)$ and is calculated as follows:
\begin{align}   
\mathcal{L}_{\beta_{\text{full}}}(N) &=  \sum_{s=1}^{n-1} \left(2^{n-s}\cdot \left \lceil \frac{2^s}{2P}  \right \rceil  \right).\label{eq:lat_sc_ps_full}
\end{align}
The decoding of a node of size $N_v$ is related to a full \gls{sc} decoding of a code of size $N$, with the final computations of \gls{ps} bits.
Hence, the execution time $\mathcal{L}_{\text{node}}(N_v)$ to decode a node of size $N_v=2^s$ is obtained using \eqref{eq:lat_sc} and \eqref{eq:lat_sc_ps_full}:  
\begin{align}
    \mathcal{L}_{\text{node}}(N_v) = \mathcal{L}_{\alpha}(N_v) + \mathcal{L}_{\beta_{\text{full}}}(N_v).
    \label{eq:lat_sc_N_full}
\end{align}

\subsection{Execution-Time Reduction (\gls{sc} Baseline Algorithm)}
\label{sec:exec_time_sav_theor}
The modified \gls{sc} trial \scmodif in \gls{grm} provides the execution-time reduction by resuming \gls{sc} from the index $\psi_{t'}$ while flipping the bits in $\bm{\varepsilon}_{t'}$. The execution time of \scmodif is denoted by $\mathcal{L}_{\scmodifmath}$ and is computed as follows:
\begin{equation}
     \mathcal{L}_{\scmodifmath} =  \mathcal{L}_{\text{SC}} - \Delta \mathcal{L}_{\text{SC}}\left(\psi_{t'}\right),
    \label{eq:lat_sc_modif}
\end{equation}
where $\Delta \mathcal{L}_{\text{SC}}\left(\psi_{t'}\right)$ denotes the execution-time reduction of \scmodif. The reduction $\Delta \mathcal{L}_{\text{SC}}\left(\psi_{t'}\right)$
provided by the \gls{grm} corresponds to the skipped \glspl{cc} 
 that would otherwise be required to calculate the \gls{llr} and \gls{ps} for the modified restart. This reduction is calculated as:
% \small
%\normalsize
%\Delta\mathcal{L}_{\text{SC}}\left(\psi_{t'}\right)  = \Delta\mathcal{L}_{\alpha}\left(\psi_{t'}\right)  + \Delta\mathcal{L}_{\beta}\left(\psi_{t'}\right) - \Theta \left(\psi_{t'}\right),
\begin{equation}   
\Delta\mathcal{L}_{\text{SC}}\left(\psi_{t'}\right)  = \Delta\mathcal{L}_{\alpha\beta}\left(\psi_{t'}\right) - \Theta \left(\psi_{t'}\right),
\label{eq:tot_saved_cc} 
\end{equation}
where $\Delta\mathcal{L}_{\alpha\beta}\left(a_0\right)$ 
denotes the skipped \glspl{cc} by avoiding \gls{llr} and \gls{ps} computations required to decode bit estimates $\{\hat{u}_0,\ldots,\hat{u}_{\psi_{t'}-1}\}$.
These two \gls{llr} and \gls{ps} components are denoted by 
$\Delta\mathcal{L}_{\alpha}\left(\psi_{t'}\right)$ and $\Delta\mathcal{L}_{\beta}\left(\psi_{t'}\right)$.
The component $\Theta\left(\psi_{t'}\right)$ corresponds to the additional \glspl{cc} required to calculate all \gls{ps} bits at the restart path of \scmodif. It is visualized with the vectors $\bm{\beta}_{\varsigma}\in \{\bm{\beta}_{3},\bm{\beta}_{1}\}$ shown in red in \autoref{fig:illustr_grm_rest_N16}.
 
 The skipped nodes inside the dashed-blue area in \autoref{fig:illustr_grm_rest_N16}  permit to visualize the reductions $\Delta\mathcal{L}_{\alpha}\left(11\right)$ and $\Delta\mathcal{L}_{\beta}\left(11\right)$. These functions are calculated as follows: 
 \begin{align}
  \Delta\mathcal{L}_{\alpha}\left(\psi_{t'}\right) &= \sum_{s=0}^{n-1} \left( \left \lfloor \frac{\psi_{t'}}{2^s} \right \rfloor \cdot \left \lceil \frac{2^s}{P}\right  \rceil\right), \label{eq:l_g_saved_cc} \\ 
 \Delta\mathcal{L}_{\beta}\left(\psi_{t'}\right) &=  \sum_{s=1}^{n-1} \left( \left \lfloor\frac{\psi_{t'}}{2^s}\right \rfloor\cdot\left \lceil \frac{2^s}{2P}  \right \rceil \right). 
 \label{eq:ps_saved_cc} 
\end{align}
In both equations, $\left \lfloor\frac{\psi_{t'}}{2^s}\right \rfloor$ indicates the number of skipped nodes at tree stage $s$.
The second term corresponds to the number of skipped \glspl{cc} to compute \gls{llr} and \gls{ps} for each node of length $N_v=2^{s}$ according to \eqref{eq:lat_sc_N_full}.
As a remainder, $P$ processing elements are used for \glspl{llr} \cite{semi_par_sc}, while $2P$ are used for \gls{ps} \cite{fast_sc}.
Note that for $s=0$, the \gls{ps} correspond to bit estimates. In this case, no computations are performed, and the summation begins from $s=1$ \eqref{eq:ps_saved_cc}.

The  component $\Theta\left(\psi_{t'}\right)$ in \eqref{eq:tot_saved_cc} is the additional \glspl{cc} for the \acrfull{ps} restoration during the restart path of \scmodif, thus being subtracted from  $\Delta\mathcal{L}_{\text{SC}}\left(\psi_{t'}\right)$. 
In this work, the time model that was proposed in \cite{fast_sc} is applied. 
In that scheme, $2P$ bits of \gls{ps} are calculated according to \eqref{eq:ps_combine} in one \gls{cc}. 
Based on this, the additional time component $\Theta\left(\psi_{t'}\right)$, expressed in \glspl{cc}, is calculated as follows: 
%Thus, $\Theta\left(\psi_{t'}\right)$ is derived as a function of $\psi_{t'}$ as follows:
\small
\begin{equation}
 \Theta\left(\psi_{t'}\right) = \sum_{s=1}^{n-1} \left(\mathcal{H}_{\psi_{t'}}\left(n-1-s\right)\cdot \left\lceil\frac{2^s}{2P}\right\rceil \cdot s \right).
\label{eq:ps_rest_cc}
\end{equation}
\normalsize

\autoref{fig:time_save_grm_all} illustrates the execution-time reduction   $\Delta\mathcal{L}_{\text{SC}}\left(\psi_{t'}\right)$  \eqref{eq:tot_saved_cc}, expressed in \glspl{cc}, for each information bit being the restart location, $\psi_{t'}=a_j \in \bm{\mathcal{A}}$, of a $\left(1024,512+11\right)$ polar code designed according to the \gls{5g} reliability order \cite{3GPP_5G_Coding}.  

\autoref{fig:tim_save_grm_p} compares the execution-time reduction induced by the \gls{grm} for two different values $P\in\{16,64\}$. Looking at \autoref{fig:tim_save_grm_p}, the reduction $\Delta\mathcal{L}_{\text{SC}}\left(\psi_{t'}\right)$ 
linearly increases as $\psi_{t'}$ moves further to the \gls{rhs} of the codeword, and it is upper bounded by the \gls{sc} reference line. For lower $P$, $\Delta\mathcal{L}_{\text{SC}}\left(\psi_{t'}\right)$ is  slightly higher for the bits further away at the \gls{rhs}. 

\autoref{fig:tim_save_grm_srm} compares execution-time reduction of \gls{grm},  \gls{srm} and \gls{lrt} for $P=64$. Looking at \autoref{fig:tim_save_grm_srm}, reduction achieved by the
 \gls{grm} is greater than that of the \gls{srm} and the \gls{lrt}, as it allows to restart from any information bit-location of the codeword. The \gls{srm} utilizes only one possible restart location $\psi_{t'}=\nicefrac{N}{2}$, resulting in a step function.
 %$\Delta\mathcal{L}_{\text{SC}}\left(\psi_{t'}\right)$ being a step function.
 The \gls{lrt} also utilizes only one possible restart location $\psi_{t'}=a_0$, resulting in a constant function.  Note that in \autoref{fig:tim_save_grm_srm}, reductions achieved by the \gls{srm} and \gls{grm} are not equal for the information bit $j_{\text{RHS}}=144$. 
 This difference occurs for two reasons. First, unlike the \gls{grm}, the \gls{srm} cannot restart from the information bit $j_{\text{RHS}}=144$, which corresponds to the location $j=a_{144}=543$. Instead, it can only restart from the frozen location $j=512$. Second, the \gls{grm} requires additional \glspl{cc} 
 for the \gls{ps} restoration $\Theta\left(543\right)$ \eqref{eq:ps_rest_cc}. Overall, the reduction of the \gls{grm} at $j_{\text{RHS}}=144$ is higher than the reduction of \gls{srm}.  
 %This is because the \gls{srm} cannot restart from this bit, and restarts only from the frozen location $j=512$. By restarting from $j_{\text{RHS}}=144$, which corresponds to the location $a_{144}=543$, the \gls{grm} is able to provide greater reduction. 

\subsection{Execution-Time Reduction ($\scLRT$ and {\fssc} Baselines)}\label{subsec:exec_time_reduc_scLRT}
\subsubsection*{$\scLRT$ Baseline Algorithm}
The execution time of $\scLRT$ trial, denoted by  $\mathcal{L}_{\scLRT}$, is calculated as \cite{Giard_JETCAS_2017}:
\begin{equation}
\mathcal{L}_{\scLRT} = \mathcal{L}_{\text{SC}}- \Delta\mathcal{L}_{\alpha\beta}\left(a_0\right),
\label{eq:lat_scLRT}
\end{equation}
%\mathcal{L}_{\scLRT} = \mathcal{L}_{\text{SC}}- \Delta\mathcal{L}_{\alpha}\left(a_0\right)- \Delta\mathcal{L}_{\beta}\left(a_0\right),
where $\Delta\mathcal{L}_{\alpha\beta}\left(a_0\right)$ 
denotes the reduction (in \glspl{cc})
by avoiding \gls{llr} and \gls{ps} computations brought by resuming \gls{sc} trial from the first information bit $a_0$. These two \gls{llr} and \gls{ps} components, denoted by 
$\Delta\mathcal{L}_{\alpha}\left(a_0\right)$ and $\Delta\mathcal{L}_{\beta}\left(a_0\right)$, are calculated according to \eqref{eq:l_g_saved_cc} and \eqref{eq:ps_saved_cc}. 
%This reduction consists of   
%and $\Delta\mathcal{L}_{\beta}\left(a_0\right)$ 
%the skipped \glspl{cc} 

The combined execution-time reduction of applying \gls{grm} to the baseline decoder $\scLRT$, denoted by $\Delta\mathcal{L}_{\scLRT}\left(\psi_{t'}\right)$, is computed as:
%\gls{lrt} $\Delta\mathcal{L}_{\scLRT}\left(\psi_{t'}\right)$ brought by %in \gls{grm} and $\scLRT$ as a baseline decoder 
 %is derived. 
%$\Delta\mathcal{L}_{\scLRT}\left(\psi_{t'}\right)$ is summarized as:
%\small
\begin{equation} 
\Delta\mathcal{L}_{\scLRT}\left(\psi_{t'}\right)  = 
 \Delta\mathcal{L}_{\text{SC}}\left(\psi_{t'}\right)- 
\Delta\mathcal{L}_{\alpha\beta}\left(a_0\right), 
 \label{eq:tot_saved_cc_scLRT} 
\end{equation}
%\begin{aligned}    
% \underbrace{\Delta\mathcal{L}_{\alpha\beta}(a_0)}_{\Delta\mathcal{L}_{\alpha}(a_0) + \Delta\mathcal{L}_{\beta}(a_0)}
%\Delta\mathcal{L}_{\alpha\beta}\left(a_0\right)
where $\Delta\mathcal{L}_{\text{SC}}\left(\psi_{t'}\right)$ is the reduction brought by the \gls{grm}, which calculated by \eqref{eq:tot_saved_cc}. 
The component $\Theta \left(\psi_{t'}\right)$ \eqref{eq:ps_rest_cc}, added by the \gls{grm}, is included in $\Delta\mathcal{L}_{\scLRT}\left(\psi_{t'}\right)$ within  $\Delta\mathcal{L}_{\text{SC}}\left(\psi_{t'}\right)$ \eqref{eq:tot_saved_cc}. 
By combining \gls{grm} and \gls{lrt} mechanisms, the total execution-time reduction will be increased. However, the reduction brought by the \gls{grm} with 
the baseline $\scLRT$ \eqref{eq:tot_saved_cc_scLRT} will be lower than that with the baseline \gls{sc} \eqref{eq:tot_saved_cc}, i.e., $\Delta\mathcal{L}_{\scLRT}\left(\psi_{t'}\right)<\Delta\mathcal{L}_{\text{SC}}\left(\psi_{t'}\right)$.

\subsubsection*{{\fssc} Baseline Algorithm}
\label{subsec:exec_time_reduc_FAST}

The following assumptions are made to estimate the execution time of the {\fsscbitflip} and {\fsscmodif} trials:
\begin{enumerate}
    \item The lengths of the special nodes $N_v$ are based on hardware implementations proposed by \cite{fast_sscf} and \cite{pract_dscf}.
    \item Decoding of a special node permits to retrieve the $N_v$ bit estimates and $N_v$ \gls{ps} 
    bits in $\left\lceil \frac{N_v}{P}\right\rceil$ \glspl{cc}.
    \item At the restart path of \fsscmodif, the \gls{ps} restorations are performed in parallel with a maximum of $2P$ operations in one \gls{cc} within each encoding stage.
    \item Outside of the special nodes, the same assumptions as for the \scmodif are kept.
\end{enumerate}

 \begin{table}[t]
    \centering    
    \normalsize
  \caption{Upper bounds of special node lengths $N_v$ \cite{pract_dscf} in {\fssc} for the Fast-\gls{scf} and Fast-\gls{dscf}-$\omega$ decoders with $\omega=\{1,2,3\}$.}
    \label{tab:special_node_size}
     \setlength{\tabcolsep}{3mm} % Default value: 6pt
     \renewcommand{\arraystretch}{1.1} % 
      %\resizebox{0.5\columnwidth}{!}{
    \begin{tabular}{cccc}%{p{0.21\columnwidth}p{0.21\columnwidth}p{0.21\columnwidth}p{0.21\columnwidth}}
        \toprule
        %\multirow{2}{*}{ } & \multicolumn{3}{c}{Upper bounds of special node lengths $N_v$}\\
         Order $\omega$ &\gls{rep} &  R1 & \gls{spc} \\
          \midrule
         1 & $N_v\leq 32$ & $N_v\leq 64$ & $N_v\leq 64$  \\
         2 & $N_v\leq 32$ & $N_v\leq 64$ & $N_v\leq 8$  \\
         3 & $N_v\leq 32$  & $N_v\leq 64$ & $N_v = 4$ \\
         \bottomrule
    \end{tabular} 
   % }    
\end{table}

\autoref{tab:special_node_size} provides upper bounds of $N_v$ for special nodes in Fast-\gls{ssc} for Fast-\gls{scf} and Fast-\gls{dscf}-$\omega$ decoders. The values of $N_v$ for the \gls{rep} and \gls{rone} nodes are selected according to implementation results of \cite{fast_sscf}, and for the \gls{spc} nodes according to implementation results of \cite{pract_dscf}. The \gls{rzero} nodes are unlimited in size and can be decoded in 1 \gls{cc} \cite{fast_sc}. 

Based on the information set and the node constraint stated in Table \ref{tab:special_node_size}, the sequence of special nodes can be found. 
This sequence is denoted by $\{n_1,\ldots, n_j,\ldots, n_q\}$, where $q$ is the total number of special nodes in the code, and $n_j$ is the $j^{\text{th}}$ node in the sequence (classified according to the order of apparition). 
%in the code denoted by 
Following that, the execution time of the {\fssc} trial, denoted by $\mathcal{L}_{\text{FSSC}}$, is estimated as: 
\begin{equation}
    \mathcal{L}_{\text{FSSC}} =  \mathcal{L}_{\text{SC}} - \Delta_{n_q},
    \label{eq:fssc_lat}
\end{equation}
where $\mathcal{L}_{\text{SC}}$ denotes the execution time of the \gls{sc} decoder \eqref{eq:lat_sc}, and $\Delta_{n_q}$ is the execution-time reduction by fast decoding the $q$ special nodes of a polar code. 

The combined execution-time reduction of \fsscmodif by restarting at $\psi_{t'}$, denoted by $\Delta \mathcal{L}_{\text{FSSC}}\left(\psi_{t'}\right)$, is calculated as:
\begin{equation}
    \Delta \mathcal{L}_{\text{FSSC}}\left(\psi_{t'}\right) = \Delta \mathcal{L}_{\text{SC}}\left(\psi_{t'} \right) - \Delta_{n_j},
    \label{eq:fssc_red_cc}
\end{equation}
where $\Delta_{n_j}$ denotes the reduction from the standard \gls{sc} by decoding the special nodes $\{n_1,\ldots,n_j\}$, with $j$ being the index of the last special node on the \gls{lhs} of $\psi_{t'}$. 
In \eqref{eq:fssc_red_cc}, additional \glspl{cc} $\Theta \left(\psi_{t'}\right)$ \eqref{eq:ps_rest_cc} induced by the \gls{grm} are included in $\Delta \mathcal{L}_{\text{SC}}\left(\psi_{t'} \right)$ \eqref{eq:tot_saved_cc}. 

By combining \gls{grm} and {\fssc}, the total execution-time reduction will be increased. However, the reduction  brought by the \gls{grm} with the baseline {\fssc} \eqref{eq:fssc_red_cc} will be lower than that with the baseline \gls{sc} \eqref{eq:tot_saved_cc}, i.e.,  $\Delta\mathcal{L}_{\text{FSSC}}\left(\psi_{t'} \right)<\Delta\mathcal{L}_{\text{SC}}\left(\psi_{t'}\right)$. 

\subsection{Memory and Partial-Sum Restoration Resources}
\label{sec:add_req_grm_mem}

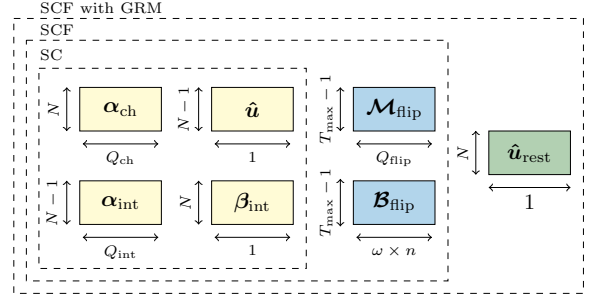
\begin{figure}[t]
\centering
%\small
\resizebox{1.0\columnwidth}{!}
{\input{Mem_struc}}
\caption{Memory sketch of \gls{sc}, \gls{scf} and \gls{scf} with \gls{grm}.}
\label{fig:mem_scf_grm}
\end{figure}

\subsubsection*{Memory}
The sketch of the memory of an \gls{scf}-based decoder with the \gls{grm} is provided in \autoref{fig:mem_scf_grm}. It is largely based on a structure we derived in our previous work \cite{simp_rest_mech} and the memory architecture proposed in \cite{fast_sscl}.
For \gls{sc} decoding, the channel \glspl{llr} $\bm{\alpha}_{\text{ch}}$ and intermediate \glspl{llr} $\bm{\alpha}_{\text{int}}$ are quantized using $Q_{\text{ch}}$ and $Q_{\text{int}}$ bits, respectively. 
The bit estimates $\bm{\hat{u}}$ are stored using $N$ bits of memory and the intermediate partial-sums $\bm{\beta}_{\text{int}}$ require with $N-1$ bits of storage. 
For \gls{scf} decoding, the metrics of the bit-flipping candidates are quantized using $Q_{\text{flip}}$ bits. Each bit-flipping set  $\bm{\varepsilon}_{t'} \in \bm{\mathcal{B}}_{\text{flip}}$  requires $\omega \times n$ bits, where $n=\log_2\left(N\right)$ is the length of the binary representation of a bit-flipping location, $\left|\bm{\mathcal{B}}_{\text{flip}}\right|=T_{\text{max}}-1$, 
and $\omega=\left|\bm{\varepsilon}_{t'}\right|_{\text{max}}$\ is the maximum size of a bit-flipping set. 

Embedding the \gls{grm} to \gls{scf}-based decoder requires additional memory in order to perform the restart path. 
It corresponds to the storage of the bit estimates $\bm{\hat{u}}_{\text{rest}}$ of initial \gls{sc} trial. 
During the restart path, the intermediate memory $\bm{\alpha}_{\text{int}}$ is restored with $\bm{\alpha}_{\varsigma}$, $\varsigma=\{n-1,\dots,0\}$, while the $\bm{\beta}_{\text{int}}$ is restored with $\bm{\beta}_{\varsigma}$. 
Thus, the memory overhead of the proposed mechanism is $N$ bits regardless of the value $\omega$. 
The total memory $\Lambda_{\text{GRM}}$ required for \gls{scf} with the \gls{grm} is then:
\begin{equation}
    \Lambda_{\text{GRM}} = \Lambda_{\text{SC}}+ \Lambda_{\text{flip}} + \Lambda_{\text{rest}}. 
    \label{eq:mem_scf_grm}
\end{equation}
 $\Lambda_{\text{GRM}}$ is composed of  memories for \gls{sc} decoding $\Lambda_{\text{SC}}$, bit-flipping candidates $\Lambda_{\text{flip}}$, and modified restart in the \gls{grm}  $\Lambda_{\text{rest}}$. The size of each of those memories is estimated according to: 
 \small
\begin{align}
    \Lambda_{\text{SC}} &= Q_{\text{ch}}\cdot N + Q_{\text{int}} \cdot (N-1) +  2N-1 \label{eq:mem_sc}, \\
    \Lambda_{\text{flip}} &= Q_{\text{flip}} \cdot \left(T_{\text{max}}-1\right) + \omega \cdot n \cdot \left(T_{\text{max}}-1\right)\label{eq:mem_scf}, \\
     \Lambda_{\text{rest}} &= N\label{eq:mem_rest}.  
\end{align}
\normalsize

\subsubsection*{Partial-Sum Restoration}
\label{sec:add_req_grm_ps_restore}
Using the architecture of \cite{fast_sc}, the encoding operations are performed with \eqref{eq:ps_combine} calculating $2P$ \gls{ps} bits in a single encoding stage $s\in\{1,\ldots,\varsigma\}$ in one \gls{cc}, where $\varsigma$ is a decoding stage where $g$-functions are performed. The total execution time required to restore all \gls{ps} vectors of interest, $\bm{\beta}_\varsigma$, is denoted by $\Theta\left(\psi_{t'}\right)$, and depends on the binary representation $\bm{\mathcal{H}}_{\psi_{t'}}$  \eqref{eq:ps_rest_cc}. 
At the restart path, a controller has to select the relevant segment of $\bm{\hat{u}}$, referred as $\bm{\eta}_{\varsigma}$ in Section \ref{sec:grm_descr}, which is on the basis for restoring $\bm{\beta}_{\varsigma}$. 
These segments are specific to each of the $k_{\text{tot}}$ possible restart locations $\psi_{t'}$, as they require the offset $\phi_{\varsigma}$ obtained by   \eqref{eq:offs_bit_est}. 
The complexity of this controller is expected to be negligible in terms of resource requirements. Also, the \gls{ps} restoration does not require additional memory as it is being stored in the intermediate partial-sums $\pmb{\beta}_{\text{int}}$. 

\section{Results and Discussions}
\label{sec:sim_res_grm}

This section compares various latency-reducing techniques listed in \autoref{tab:sum_dif_schemes}.
%are compared. 
The execution-time reduction provided by the proposed \gls{grm} is estimated
%by simulations while 
by applying \eqref{eq:tot_saved_cc}, \eqref{eq:tot_saved_cc_scLRT}, and \eqref{eq:fssc_red_cc}. Finally, the applicability of \gls{grm} to other flip-based polar decoders is discussed.

\subsection{Simulation Setup}
The effects of the \gls{grm} on \gls{scf} and \gls{dscf}-$\omega$ decoders are observed through the simulations. 
The \gls{5g} polar codes \cite{3GPP_5G_Coding} of length $N=1024$ are used with three different code rates $R\in\{\nicefrac{1}{8},\nicefrac{1}{4},\nicefrac{1}{2}\}$. The \gls{5g} polynomial $z^{11}+z^{10}+z^{9}+z^{5}+1$ generates  $r=11$ \gls{crc} codes. 
The number of processing elements is set to $P=64$ as in \cite{Giard_JETCAS_2017,pract_dscf}.
The \gls{bpsk} modulation is used over an \gls{awgn} channel. Simulations are performed by estimating a minimum of $C=2\cdot 10^5$ random codewords or until $2000$ frame errors are observed.

%The values of the maximum number of trials $T_{\text{max}}$ are selected as described in Subsection~\ref{sec:sim_res_sc_dscf_intro} of Chapter~\ref{chapt:flip_dec}.
The maximum number of trials $T_{\text{max}}$ were selected to achieve an error-correction performance that is close to the genie-aided decoder \cite{dyn_scf} at the target \gls{fer} of $10^{-2}$.
For the single bit-flip \gls{scf} and \gls{dscf}-1 decoders, $T_{\text{max}}\in\{13,8\}$ is set, and for the multi-bit flip \gls{dscf}-$\omega$ decoders with $\omega\in\{2,3\}$, $T_{\text{max}}\in\{51,301\}$ is set. 
%For the \gls{dscf}-$\omega$ decoders, selection of $T_{\text{max}}$ is motivated by achieving the error-correction performance that is close to the genie-aided decoders for $\omega\in\{1,2,3\}$ at the target \gls{fer} of $10^{-2}$ \citep{dyn_scf}. 
%For the \gls{scf} decoder, $\Tmax=13$ is set, which is close to $\Tmax=8$ in the \gls{dscf}-1 decoder. 
The hardware-friendly metric calculation function \eqref{eq:metr_dscf_w_simp} is applied for all \gls{dscf}-$\omega$ decoders. The comparisons are made in terms of the error-correction performance, memory requirements, and average execution time. 
The hardware-friendly metric calculation function \eqref{eq:metr_dscf_w_simp} is applied as part of the metric calculations to all \gls{dscf} decoders. 
For the Fast-\gls{scf} and Fast-\gls{dscf}-$\omega$ decoders, 
the upper bounds for special node lengths are selected according to \autoref{tab:special_node_size}.

\subsection{Error-Correction Performance}

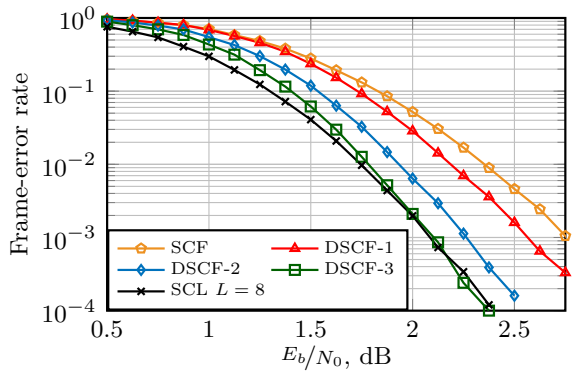
\begin{figure}[t]
\centering
\resizebox{1.0\columnwidth}{!}{
\input{FER_SCF_SCL}}
\caption{Error-correction performance of \gls{scf}, \gls{dscf}-$\omega$ and \gls{scl} decoders for a $(1024,512+11)$ code.}
\label{fig:fer_scf_scl}
\end{figure}

\autoref{fig:fer_scf_scl} depicts the error-correction performance for a $(1024,512+11)$ \gls{5g} polar code with the \gls{scf}, \gls{dscf}-1, \gls{dscf}-2, and \gls{dscf}-3 decoders. The $y$-axis corresponds to the \gls{fer}, while the $x$-axis represents the $\nicefrac{E_b}{N_0}$ (energy per information bit to the noise spectral density). 
The performance of the \gls{scl} decoder with $L=8$  represents the baseline 
for the \gls{5g} performance evaluation \cite{3GPP_5G_Coding}. The results show that the \gls{dscf}-3 decoder closely approaches the error-correction performance of the \gls{scl} decoder with $L=8$ with only around $0.05$ dB loss at \gls{fer} of $10^{-2}$. 
Since the \gls{5g} polar codes are the targeted codes in this paper and according to the performance, \gls{dscf}-3 is the algorithm of reference in the remaining part of this section. Our proposed \gls{grm} provides a greater execution-time reduction for this decoder. 

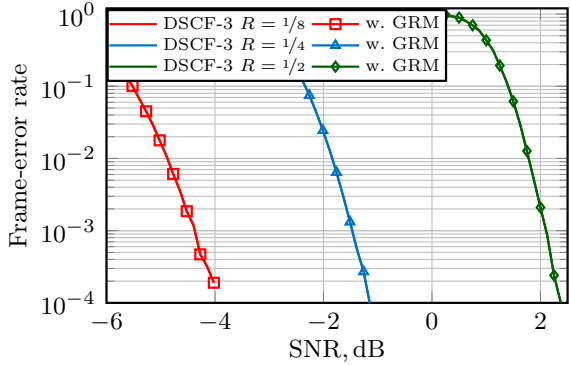
\begin{figure}[t]
\centering\resizebox{1.0\columnwidth}{!}{
\input{FER_DSCFw}}
\caption{Error-correction performance for $N=1024$ code of a \gls{dscf}-3 decoder with and without the \gls{grm}.}
\label{fig:fer_dscfw}
\end{figure}

\autoref{fig:fer_dscfw} depicts the error-correction performance  for the \gls{dscf}-3 decoder with and without the \gls{grm} for the code with various rates. 
%for polar codes of $N=1024$ and $R=\{\nicefrac{1}{8},\nicefrac{1}{4},\nicefrac{1}{2}\}$.  The $y$-axis corresponds to the \gls{fer}, while the $x$-axis represents the \gls{snr}.  
The results show that the \gls{grm} has no detrimental effect on the error-correction performance. 
This is expected and in line with the definition of the mechanism provided in Section~\ref{sec:grm_descr}.
The \gls{lrt} does not affect the error-correction performance. The Fast-\gls{scf} and Fast-\gls{dscf}-$\omega$ decoders have  negligible degradation of the error correction \cite{fast_sscf,scf_eff_impl,pract_dscf}. 

\subsection{Memory Estimations}
Memory requirements are estimated with \eqref{eq:mem_scf_grm}, where the same quantization scheme as that of \cite{pract_dscf,simp_rest_mech} is used. The blocks based on the \gls{llr} values are quantized by $Q_{\text{ch}}=6$, $Q_{\text{int}}=7$, and $Q_{\text{flip}}=7$ bits, respectively. 
The sizes of these blocks also depend on the values of $N$ and $T_{\text{max}}$, but are independent from $R$. 
The memory estimates in bits and the memory overhead are provided in \autoref{tab:mem_scf_grm}. 
The estimations show that the proposed \gls{grm} leads to a memory overhead of $3.87\%$ to $6.58\%$.  
The smallest memory overhead is obtained for the \gls{dscf}-3 algorithm. This happens because the \gls{dscf}-3 decoder requires the largest memory due to a higher value of $T_{\text{max}}$, which reduces the impact of the \gls{grm}.  
The memory overhead is the same when flip decoders are used with other baseline algorithms (\gls{lrt} and {\fssc}).
Therefore, the overhead of $3.87\%$ to $6.58\%$ is considered for \gls{grm} regardless of the baseline algorithm. 

\begin{table}[t]
\centering
\scriptsize
%\parbox{0.95\textwidth}{
\caption{Memory estimates and overhead for flip decoders with and without the \gls{grm} for a $N=1024$ code.}
\label{tab:mem_scf_grm}%}
\setlength{\tabcolsep}{3pt} % Default value: 6pt
\renewcommand{\arraystretch}{1.3} % Default value: 1
% \resizebox{1.0\columnwidth}{!}{
\begin{tabular}{ccccc}
\toprule
\multirow{2}{*}{Decoders} & \multirow{2}{*}{$T_{\text{max}}$} & no \gls{grm} & w. \gls{grm}  & Mem. overh. \\ 
& & bits & bits & $\%$ \\
\midrule
\gls{scf} & $13$ & $15556$  & $16580$  & $6.58$  \\ %\hline
\gls{dscf}-1 & $8$ & $15471$  & $16495$  & $6.62$  \\ %\hline 
\gls{dscf}-2 & $51$ & $16702$ & $17726$ & $6.13$   \\ %\hline 
\bf{\gls{dscf}-3} &  $\mathbf{301}$ & $\mathbf{26452}$ & $\mathbf{27476}$ & $\mathbf{3.87}$  \\ %\hline
\bottomrule
\end{tabular}
%}
\end{table}

\subsection{Average Execution Time}
\begin{table}[t]
\centering
\tiny
\caption{Reduction $\avgreducgrm$ \eqref{eq:lat_sav_grm} by applying the \gls{grm} to flip decoders with various baselines $\dec$ at the \gls{fer}~$10^{-2}$.} 
\label{tab:gain_fer_grm_all}
%\review{Codes are designed according to \gls{5g}\cite{3GPP_5G_Coding}.}}
\setlength{\tabcolsep}{1.0pt} % Default value: 6pt
\renewcommand{\arraystretch}{1.1} % Default value: 1
% \resizebox{1.0\columnwidth}{!}{
\begin{tabular}{ccccccccc}
\toprule
& & & 
\multicolumn{2}{c}{$R=\nicefrac{1}{8}$} & \multicolumn{2}{c}{$R=\nicefrac{1}{4}$} & \multicolumn{2}{c}{$R=\nicefrac{1}{2}$} \\ 
\midrule
\multirow{2}{*}{dec} &\multirow{2}{*}{flip} & \multirow{2}{*}{$T_{\text{max}}$} &  $\nicefrac{E_b}{N_0}$ & $\avgreducgrm$ & $\nicefrac{E_b}{N_0}$ & $\avgreducgrm$ & $\nicefrac{E_b}{N_0}$ & $\avgreducgrm$  \\ %\hline 
 & & & dB & $\%$ & dB & $\%$ & dB & $\%$ \\ %\hline
 \midrule
 %\gls{scf} & $13$ & $2.00$ & $19.69$ & $1.875$ & $15.39$ & $2.375$ & $10.87$ \\ 
 \multirow{4}{*}{\gls{sc}}& \gls{scf} & $13$ & $2.00$ & $15.81$ & $1.75$  &  $18.06$ & $2.375$ & $10.50$ \\ 
 & \gls{dscf}-1 &  $8$ &  $1.75$ &  $12.27$  &  $1.625$  &  $10.81$  &  $2.25$ &  $5.00$\\ 
& \gls{dscf}-2 &  $51$ &   $1.375$ &  $38.00$  &  $1.375$ &   $29.46$ &  $2.00$ &  $15.71$ \\ 
& \bf{\gls{dscf}-3} &  \bf{ $\mathbf{301}$} &  $\mathbf{1.125}$ &   $\mathbf{56.90}$ &   $\mathbf{1.125}$ &  $\mathbf{46.18}$ &  $\mathbf{1.75}$  &  $\mathbf{26.00}$  \\ 
 \midrule
%\midrule
\multirow{4}{*}{ $\scLRT$}& \gls{scf} &  $13$ &  $2.00$ &  $13.22$ &  $1.75$ &  $16.24$ &  $2.375$ &  $9.50$\\
& \gls{dscf}-1 &  $8$ &  $1.75$ &  $8.53$ &  $1.625$ &  $8.76$ &  $2.25$ & $4.03$ \\
& \gls{dscf}-2 & $51$ & $1.375$ & $24.09$ & $1.375$ & $22.61$ & $2.00$ & $11.81$\\
&\bf{\gls{dscf}-3} & $\mathbf{301}$ & $\mathbf{1.125}$ & $\mathbf{33.09}$ & $\mathbf{1.125}$ & $\mathbf{33.32}$ & $\mathbf{1.75}$ & $\mathbf{17.83}$ \\
 \midrule
\multirow{4}{*}{FSSC}&\gls{scf} & $13$ & $2.00$ & $10.27$  & $1.75$ & $14.67$  & $2.375$ & $10.60$ \\
&\gls{dscf}-1 & $8$ & $1.75$ &  $5.90$ & $1.625$ & $7.10$ & $2.25$ & $4.24$ \\
&\gls{dscf}-2 & $51$ & $1.375$ &  $16.03$ & $1.375$ & $16.24$  & $2.00$ & $11.10$ \\
&\bf{\gls{dscf}-3} & $\mathbf{301}$ & $\mathbf{1.125}$ & $\mathbf{20.87}$  & $\mathbf{1.125}$ &  $\mathbf{22.14}$ & $\mathbf{1.75}$ & $\mathbf{15.21}$ \\
\bottomrule
\end{tabular}%}
\end{table}

\begin{table}[t]
\centering
\footnotesize
\caption{Reduction $\Delta\overline{\mathcal{L}}_{\text{flip}}$  \eqref{eq:lat_sav_grm} by applying different mechanisms to \gls{dscf}-3 decoder at the \gls{fer}~$10^{-2}$.} %\review{Codes are designed according to \gls{5g} \cite{3GPP_5G_Coding}.}}
\setlength{\tabcolsep}{1.5pt} % Default value: 6pt
\renewcommand{\arraystretch}{1.1} % Default value: 1
\begin{tabular}{cccccccc}
\toprule
 &  &
\multicolumn{2}{c}{$R=\nicefrac{1}{8}$} & \multicolumn{2}{c}{$R=\nicefrac{1}{4}$} & \multicolumn{2}{c}{$R=\nicefrac{1}{2}$} \\ 
\midrule
\multirow{2}{*}{tech.} & \multirow{2}{*}{$T_{\text{max}}$} &  $\nicefrac{E_b}{N_0}$ & $\Delta\overline{\mathcal{L}}_{\text{flip}}$ & $\nicefrac{E_b}{N_0}$ & $\Delta\overline{\mathcal{L}}_{\text{flip}}$ & $\nicefrac{E_b}{N_0}$ & $\Delta\overline{\mathcal{L}}_{\text{flip}}$  \\  
 &   &dB & $\%$ & dB & $\%$ & dB & $\%$ \\
 \midrule 
 \bf{w. \gls{grm}} &  \multirow{3}{*}{$301$}  &  \multirow{3}{*}{$1.125$}  & $\mathbf{56.90}$  & \multirow{3}{*}{$1.125$} & $\mathbf{46.18}$ &  \multirow{3}{*}{$1.75$} & $\mathbf{26.00}$ \\
 w. \gls{srm}~\cite{simp_rest_mech} & &  &  $30.05$  & 
 & $17.90$ & & $4.04$ \\ 
 w. LRT~\cite{Giard_JETCAS_2017} &   &   & $46.08$  &  & $24.20$ & & $11.84$ \\
%\hline
%\bf{w. \gls{grm}} &  &  &  $\mathbf{59.72}$ &  & $\mathbf{46.44}$ &  & $\mathbf{28.55}$ \\ %\hline 
\bottomrule
\end{tabular}
\label{tab:gain_fer_dscf3_srm_grm}
\end{table}

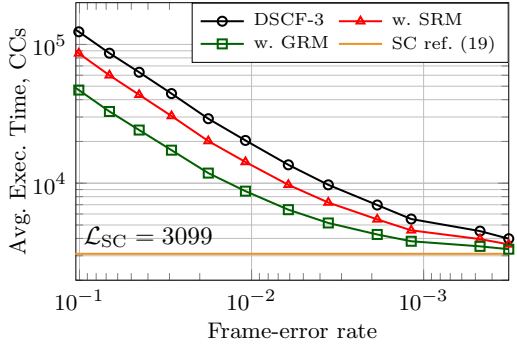
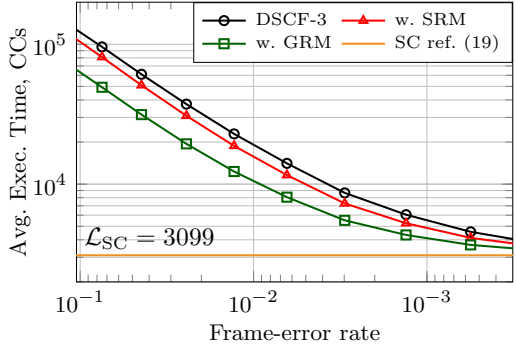
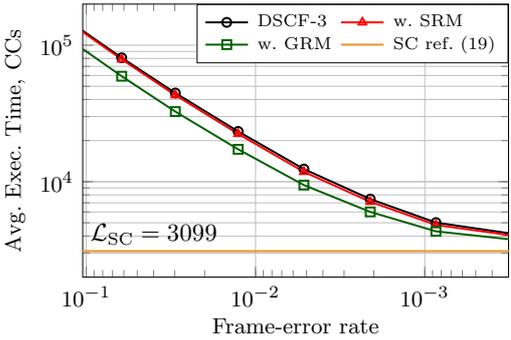
\begin{figure}[ht]
 \centering	
 \begin{subfigure}[t]{1.0\columnwidth}
  \centering
  \resizebox{0.9\columnwidth}{!}{	  \input{Aver_exec_time_GRM_SRM_R0125.tex}}
  %\input{Aver_exec_time_GRM_SRM_R0125.tex}
  %\vspace{-0.2cm}
  \caption{$R=\nicefrac{1}{8}$.}
 % \vspace{0.1cm} 
	\label{fig:av_exec_dscf_grm_r_0125}
 \end{subfigure}
 \begin{subfigure}[t]{1.0\columnwidth}
 \centering\resizebox{0.9\columnwidth}{!}{	 
 \input{Aver_exec_time_GRM_SRM_R025.tex}}
 %}
% \vspace{-0.2cm} 
 \caption{$R=\nicefrac{1}{4}$.}
 %\vspace{0.1cm} 
  \label{fig:av_exec_dscf_grm_r_025}
 \end{subfigure}
 \begin{subfigure}[t]{1.0\columnwidth}
  \centering
  \resizebox{0.9\columnwidth}{!}{
  \input{Aver_exec_time_GRM_SRM_R05.tex}}
  \vspace{-0.2cm}
  \caption{$R=\nicefrac{1}{2}$.}  
   \label{fig:av_exec_dscf_grm_r_05}
 \end{subfigure} 
  \caption{Average execution time of \gls{dscf}-3 with the \gls{grm}, \gls{srm} and the original decoder for $R=\{\nicefrac{1}{8},\nicefrac{1}{4},\nicefrac{1}{2}\}$.} 
 \label{fig:av_exec_dscf_grm_r_all}
\end{figure}

\begin{figure}[ht]
 \centering	   
  \resizebox{1.0\columnwidth}{!}{\input{Aver_exec_time_all_R025}}
  \caption{Average execution time of flip decoders with and without the proposed \gls{grm} for all $\dec\in\{\text{SC},\scLRT,\text{\fssc}\}$ for a $\left(1024,256+11\right)$ 5G polar code.}   
 \label{fig:av_exec_dscf_lrt_fast_grm_r025}
\end{figure}
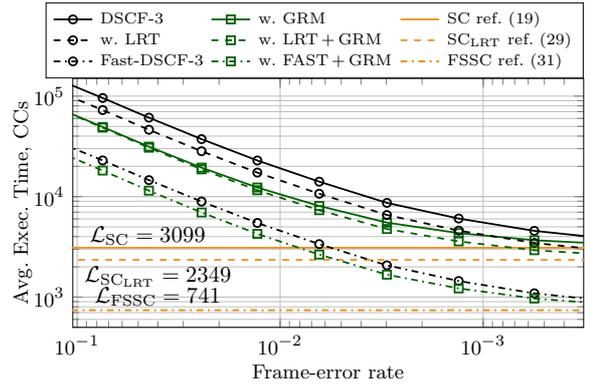

% Also, the delay associated with updating the bit-flipping sets for the \gls{dscf}-$\omega$ decoders is not considered in the time model of this thesis, as explained in Section~\ref{sec:exec_time_mod}. Therefore, execution times of an additional trial of the \gls{scf} and \gls{dscf} decoders are considered equivalent, and both are referred to as \emph{flip decoders}. 

For \gls{dscf}-$\omega$ decoders with $\omega>1$,
 the delay associated with updating the bit-flipping sets for the \gls{dscf}-$\omega$ decoders is not considered in this work, 
 %the delay associated with the updates of the bit-flipping sets is not considered in the time model of this work, 
 since it was shown to have a low impact on overall execution time compared to \gls{sc} decoding tree traversal \cite{pract_dscf}.  
Therefore, the following equations are considered equivalent for \gls{scf} and \gls{dscf}-$\omega$ decoders, and both decoders are referred to as  ``$\flip$'' decoders. 
%Next, we refer as ``$\flip$'' decoder, either \gls{scf} or \gls{dscf}-$\omega$.

The execution time of the flip decoder 
to decode the  $c^{\text{th}}$ codeword is denoted by $l_{\flip}\left(c\right)$. It is 
expressed in 
\glspl{cc} and calculated as follows:
%calculated as the product 
\begin{equation}
l_{\flip}\left(c\right) = \tau\left(c\right) \cdot \mathcal{L}_{\dec},
\label{eq:lat_cw_scf}
\end{equation}
where $\tau\left(c\right)$ denotes the required number of trials, including the initial trial, and  $\mathcal{L}_{\dec}$ denotes the baseline algorithm latency, where $\dec\in\{\text{SC},\,\scLRT,\,\text{\fssc}\}$. These baseline latencies are computed by \eqref{eq:lat_sc}, \eqref{eq:lat_scLRT}, and \eqref{eq:fssc_lat}. 

By applying the \gls{grm}, the execution time to estimate the $c^{\text{th}}$ codeword is modified as follows:
%Next, we denote by $l_{\flipgrm}(c)$ the execution time to decode a codeword $c$ by the flip decoder with the proposed \gls{grm}. It is obtained as follows:
\begin{equation}
 l_{\flipgrm}\left(c\right) = l_{\flip}\left(c\right)-\sum_{t'=1}^{\tau'\left(c\right)} \Delta\mathcal{L}_{\dec}\left(\psi_{t'}\right),
\label{eq:lat_mod_sc_scf}
\end{equation}
where $\Delta\mathcal{L}_{\dec}\left(\psi_{t'}\right)$ refers to  \eqref{eq:tot_saved_cc},  \eqref{eq:tot_saved_cc_scLRT}, \eqref{eq:fssc_red_cc} depending on the baseline algorithm $\dec$, and $\tau'(c)$ is the required number of additional trials for this codeword. 
The average execution times of flip decoders with and without the \gls{grm} are denoted $\overline{\mathcal{L}}_{\flipgrm}$ and $\overline{\mathcal{L}}_{\flip}$.
Both are calculated using a total of $C$ estimated codewords as follows:
\begin{equation}    
\begin{aligned}
\overline{\mathcal{L}}_{\flip} &= \frac{1}{C}\sum_{c=0}^{C-1} l_{\flip}\left(c\right), \\
\overline{\mathcal{L}}_{\flipgrm} &= \frac{1}{C}\sum_{c=0}^{C-1} l_{\flipgrm}\left(c\right).
\label{eq:lat_scf_grm}
\end{aligned}
\end{equation}

The reduction of the average execution time of a  flip decoder with the \gls{grm} compared to the original flip decoder, denoted by 
 $\Delta\overline{\mathcal{L}}_{\flip}$, is computed as: 
%The average execution-time reduction of a flip decoder with the \gls{grm} with respect to the original flip decoder is denoted % $\Delta\overline{\mathcal{L}}_{\flip}$ and estimated by simulation as:  
\begin{equation}
\avgreducgrm = \frac{\overline{\mathcal{L}}_{\flip}-\overline{\mathcal{L}}_{\flipgrm}}{\overline{\mathcal{L}}_{\flip}}.
\label{eq:lat_sav_grm}
\end{equation}

\autoref{fig:av_exec_dscf_grm_r_all} 
shows the average execution time of a \gls{dscf}-3 decoder for rates $R=\{\nicefrac{1}{8},\nicefrac{1}{4},\nicefrac{1}{2}\}$, respectively. 
In these results, the original \gls{sc} is used as a baseline algorithm. 
The presented algorithms are \gls{dscf}-3, \gls{dscf}-3 with the \gls{grm}, and \gls{dscf}-3 with the \gls{srm} \cite{simp_rest_mech}.
Additionally, the execution time of \gls{sc} decoding is shown for reference, which corresponds to the bound of these decoders. 
%The execution time of \gls{sc} decoding is shown for a reference and corresponds to the bound of these algorithms. 
The results show that the decoder with the \gls{grm} achieves significant reduction for all code rates and throughout the practical range of the \gls{fer}. 

\autoref{fig:av_exec_dscf_lrt_fast_grm_r025} shows the average execution time of a \gls{dscf}-3 decoder for $R=\nicefrac{1}{4}$. In these results, all $\dec\in\{\text{SC},\scLRT,\text{\fssc}\}$ are applied as baseline algorithms, and their latencies are also shown for reference. Comparing the results for \gls{grm} and sole \gls{lrt}, \gls{grm} shows a greater reduction. However, as the \gls{fer} decreases, \gls{lrt} surpasses the \gls{grm} and even \gls{sc}. 
This happens due to the differences in baseline algorithms.
The $\scLRT$ is the baseline algorithm in \gls{dscf}-3 with the \gls{lrt}, whereas the original \gls{sc} is the baseline in \gls{dscf}-3 with the \gls{grm}. This difference is more noticeable at low \gls{fer}, where the initial trials tend to prevail on average.  
% the combined mechanism, \gls{lrt}+\gls{grm}, achieves a lower average execution time compared to other mechanisms. Notably, for a low-rate code of $R=\nicefrac{1}{8}$, the \gls{lrt}+\gls{grm} has a higher reduction at the target \gls{fer} of $10^{-2}$, compared to  other mechanisms. However, for rates $R\in\{\nicefrac{1}{4}, \nicefrac{1}{2}\}$, the difference between this combined mechanism and the sole \gls{grm} is negligible at this \gls{fer}.
% Notably, for a low-rate code of $R=\nicefrac{1}{8}$, the \gls{lrt}+\gls{grm} has a higher reduction at the target \gls{fer} of $10^{-2}$, compared to  other mechanisms.
% However, for rates $R\in\{\nicefrac{1}{4}, \nicefrac{1}{2}\}$, the difference between this combined mechanism and the sole \gls{grm} at this \gls{fer} is negligible.
The average execution time of the combined mechanism  \gls{lrt}+\gls{grm} achieves a lower average execution time compared to each sole mechanism. 
However, at the \gls{fer} of $10^{-2}$, the difference between \gls{lrt}+\gls{grm} and a sole \gls{grm} is negligible.
Comparing the results for Fast-\gls{dscf}-3, the \gls{grm} achieves a noticeable reduction of the average execution time compared to the standard Fast-\gls{dscf}-3 decoder. This reduction remains significant throughout the practical \gls{fer} range of $10^{-2}$ to $10^{-3}$. The upper bounds for special node lengths are selected according to line $\omega=3$ in \autoref{tab:special_node_size}. 

%Table~\ref{tab:gain_fer_dscf3_srm_grm} summarizes reductions for \gls{dscf}-3 decoder with the \gls{grm} and \gls{srm}. Compared to original decoding, the \gls{dscf}-3 with the \gls{grm} achieves the reductions in $\Delta\overline{\mathcal{L}}_{\flip}$ of $\{56.90\%, 46.18\%,26.00\%\}$, while for \gls{srm}, it achieves $\{30.05\%,17.90\%,4.04\%\}$. Comparing two mechanisms, 
% the \gls{grm} provides higher reductions for all code rates. The \gls{grm} requires the $N$ bits of additional memory.

The numerical results of the average execution-time reduction with respect to original decoding $\Delta\overline{\mathcal{L}}_{\flip}$, estimated by \eqref{eq:lat_sav_grm} at the target \gls{fer} of $10^{-2}$ in percent, are provided in  \autoref{tab:gain_fer_grm_all}
and \autoref{tab:gain_fer_dscf3_srm_grm}.
 \autoref{tab:gain_fer_grm_all} provides results of applying the \gls{grm} to flip decoders using different baseline algorithms.
 %and  Table~\ref{tab:gain_fer_dscf3_srm_grm} summarizes reductions for \gls{dscf}-3 decoder with the \gls{grm} and \gls{srm}. 
%In \autoref{tab:gain_fer_grm_all}, reductions are provided for flip decoders with \gls{grm} using different baseline algorithms. 
Compared to original decoding, \gls{dscf}-3 with the \gls{grm} and with \gls{sc} as a baseline achieves the reductions $\Delta\overline{\mathcal{L}}_{\flip}$ of $\{56.90\%, 46.18\%,26.00\%\}$. The \gls{dscf}-3 with \gls{grm} and $\scLRT$ as a baseline achieves the reductions $\Delta\overline{\mathcal{L}}_{\flip}$ of $\{33.09\%, 33.32\%,17.83\%\}$. The overall reductions of \gls{dscf}-3 with \gls{lrt}+\gls{grm} compared to the original \gls{dscf}-3 decoder are $\{64\%,50\%,28\%\}$.  
 The \gls{dscf}-3 with the \gls{grm} and {\fssc} as a baseline algorithm achieves the reductions $\Delta\overline{\mathcal{L}}_{\flip}$ of $\{20.87\%, 22.14\%, 15.21\%\}$. The overall reductions of \gls{dscf}-3 with {\fssc}+\gls{grm} are approximately $\{86\%, 81\%, 74\%\}$.
 The reductions for \gls{dscf}-3 decoder with the \gls{grm}, \gls{srm}, and \gls{lrt} compared to the original \gls{dscf}-3 are provided in \autoref{tab:gain_fer_dscf3_srm_grm}. 
These results indicate that the \gls{grm} achieves higher reductions compared to \gls{srm} and \gls{lrt}. 

According to the results of \autoref{tab:gain_fer_grm_all} and  \autoref{tab:gain_fer_dscf3_srm_grm}, the higher reductions are achieved for the flip decoders with \gls{sc} as the baseline algorithm, while the lower reductions are observed for those with {\fssc} as the baseline. This is expected, as \gls{grm} is an additional mechanism applied to the fast decoder, which already provides a reduction compared to the original decoding.
Nevertheless, the reduction provided by the \gls{grm} is significant. 
The proposed \gls{grm} requires $N$ bits of additional memory for all flip decoders with any baseline algorithm.
%As expected in Section \ref{subsec:exec_time_reduc_scLRT}, the greatest reduction is achieved for flip decoders with \gls{sc} as a baseline algorithm, and the smallest reduction is achieved for flip decoders with {\fssc}.
%Nevertheless, this reduction is significant. 
%The proposed \gls{grm} requires $N$ bits of additional memory for all flip decoders, which is equal for all baseline algorithms. 
%$\psi_{t'}$ 

\subsection{Applicability to Other Flip Decoding Algorithms}
\label{sec:other_scf}
The proposed \gls{grm} rests upon the sequential nature of \gls{sc} to provide execution-time reduction for \gls{scf}-based decoding.
The \gls{scan} \cite{scan_intro} can be viewed as a tree decoding algorithm of \gls{sc} with \glspl{llr} being traversed the tree both ways. Hence, \gls{scan} flip \cite{scanf,eff_scanf_hw} decoder will benefit from restarting at the flip location 
rather than $0$. However, restarting from a specific point in the decoding tree will require storing more than just the $N$ bits $\bm{\hat{u}}_{\text{rest}}$, as \glspl{llr} must also be stored, particularly when internal soft iterations are used. The \gls{bp} has a parallel decoding schedule, and tree traversal is not used as in \gls{sc}-based decoders. Therefore, the restart from a certain bit in the codeword does not benefit \gls{bpf}, compared to \gls{scf} decoding. 

\glsreset{grm}
\section{Conclusion}
\label{sec:conclusion}

In this work, we proposed the \gls{grm} to reduce the average execution time of \gls{scf}-based decoders by skipping parts of the decoding tree to estimate the bit-flipping candidate and all the previous bits in each additional decoding trial. The decoding tree is traversed from the root along the restart path to directly estimate the restart bit. To perform such a restart, the \acrfull{ps} bits are restored from the bit estimates stored in memory following the initial unsuccessful decoding trial. 
These \gls{ps} restorations are shown to be made by low-complexity encoding operations. The proposed \gls{grm} can be used with any flavor of \gls{scf} decoding and does not affect the error-correction performance. The \gls{grm} can be effectively combined with state-of-the-art fast decoding techniques. When applied to the \gls{dscf}-3 decoder, the proposed \gls{grm} reduced the average execution time by $26\%$ to $60\%$ for the $N=1024$ polar code.
In addition to the improvements achieved by fast decoding, applying the \gls{grm} to the Fast-\gls{dscf}-3 decoder further reduced the average execution time by $15\%$ to $22\%$. Across all \gls{dscf}-3 decoders, the \gls{grm} was found to increase memory requirements by approximately $4\%$.

%When applied to the Fast-\gls{dscf}-3 decoder, the \gls{grm} reduced the average execution time by $15\%$ to $22\%$, in addition to the improvements already achieved by fast decoding.  For all \gls{dscf}-3 decoders, the \gls{grm} resulted in approximately $4\%$ of additional memory.

%When applied to the \gls{dscf}-3 with the \gls{lrt}, the \gls{grm} reduced the average execution time by $17\%$ to $33\%$, in addition to the improvements already achieved by the \gls{lrt}. 

% IEEEabrv,ConfAbrv,
\bibliography{IEEEabrv,ConfAbrv,bibliography}
%\input sn-article.bbl

\end{document}

%% file: sc_tree.tex
\begin{tikzpicture}[]
\node [draw, circle, line width=0.8pt, inner sep=0.1 cm, label=south:{$\hat{u}_0$}] (u0) at (0,0){};
\node [draw, circle, line width=0.8pt, inner sep=0.1 cm, label=south:{$\hat{u}_1$}] (u1) at (0.6,0){};
\node [draw, circle, line width=0.8pt, inner sep=0.1 cm, label=south:{$\hat{u}_2$}] (u2) at (1.2,0){};
\node [draw, circle, line width=0.8pt, inner sep=0.1 cm, label=south:{$\hat{u}_3$}, fill=black] (u3) at (1.8,0){};
\node [draw, circle, line width=0.8pt, inner sep=0.1 cm, label=south:{$\hat{u}_4$}] (u4) at (2.4,0){};
\node [draw, circle, line width=0.8pt, inner sep=0.1 cm, label=south:{$\hat{u}_5$}, fill=black] (u5) at (3.0,0){};
\node [draw, circle, line width=0.8pt, inner sep=0.1 cm, label=south:{$\hat{u}_6$}, fill=black] (u6) at (3.6,0){};
\node [draw, circle, line width=0.8pt, inner sep=0.1 cm, label=south:{$\hat{u}_7$}, fill=black] (u7) at (4.2,0){};

\node [draw, circle, line width=0.8pt, inner sep=0.1 cm, fill=gray!50!white] (n10) at (0.3,0.5){};
\node [draw, circle, line width=0.8pt, inner sep=0.1 cm, fill=gray!50!white] (n11) at (1.5,0.5){};
\node [draw, circle, line width=0.8pt, inner sep=0.1 cm, fill=gray!50!white] (n12) at (2.7,0.5){};
\node [draw, circle, line width=0.8pt, inner sep=0.1 cm, fill=gray!50!white] (n13) at (3.9,0.5){};
\node [draw, circle, line width=0.8pt, inner sep=0.1 cm, fill=gray!50!white] (n20) at (0.9,1.0){};
\node [draw, circle, color=red,  line width=0.8pt, inner sep=0.05 cm, fill=gray!50!white] (n21) at (3.3,1.0){\textcolor{black}{\footnotesize{$v$}}};
\node [draw, circle, line width=0.8pt, inner sep=0.1 cm, outer sep=0.01 cm,   fill=gray!50!white] (n30) at (2.1,1.5){};

\node[inner sep=0] (s00) at (-0.7, 1.9) {\small{$s$}};
\draw ($(s00.south)+(-0.1,-0.07)$)--($(s00.south)+(0.1,-0.07)$);

\node[inner sep=2] (s3) at (-0.7, 1.5) {\footnotesize{$3$}};
\node[inner sep=2] (s2) at (-0.7, 1.0) {\footnotesize{$2$}};
\node[inner sep=2] (s1) at (-0.7, 0.5) {\footnotesize{$1$}};
\node[inner sep=2] (s0) at (-0.7, 0.00) {\footnotesize{$0$}};
 \draw[dotted] (s3)--(n30);
 \draw[dotted] (s2)--(n20);
 \draw[dotted] (s1)--(n10);
 \draw[dotted] (s0)--(u0);
\draw[blue] (u0) -- node{}(n10);
\draw[red] (u1) -- node{}(n10);
\draw[blue] (u2) -- node{}(n11);
\draw[red] (u3) -- node{}(n11);
\draw[blue] (u4) -- node{}(n12);
\draw[red] (u5) -- node{}(n12);
\draw[blue] (u6) -- node{}(n13);
\draw[red] (u7) -- node{}(n13);
\draw [line width=0.8pt,blue] (n10) -- node{}(n20);
\draw [line width=0.8pt,red] (n11) -- node{}(n20);
\draw [line width=0.8pt,blue] (n12) -- node{}(n21);
\draw [line width=0.8pt,red] (n13) -- node{}(n21);
\draw [line width=1.2pt,blue] (n20) -- node{}(n30);
\draw [line width=1.2pt,red] (n21) -- node{}(n30);
\draw [-latex] (2.3,1.56) -- node[anchor=south] {\footnotesize{$\bm{\alpha}_{v}$}}(3.2,1.2);
\draw [latex-] (2.2,1.36) -- (3.1,1.0);
\draw (2.3,1.1) node {\footnotesize{$\bm{\beta}_{v}$}};

%\draw [dashed, -latex] (3.0,1.0) -- node[anchor=north east] {\footnotesize{$\beta_v$}}(2.2,1.33);

% in and out from v node, left branches
\draw [-latex] (3.1,0.95) -- node{} (2.75,0.68);
\draw [-latex] (2.88,0.53) -- node{} (3.22,0.8);

% right branches
\draw [latex-] (3.85,0.68) -- node{} (3.5,0.95);
\draw [latex-] (3.36,0.82) -- node{} (3.7,0.53);

\node [] (al) at (2.65,0.95){\footnotesize{$\bm{\alpha}_v^l$}};

\node [] (bl) at (3.1,0.45){\footnotesize{$\bm{\beta}^l$}};

%\node [] (ar) at (3.47,0.37){\footnotesize{$\bm{\alpha}_v^r$}};

%\node [] (br) at (3.86,0.96){\footnotesize{$\bm{\beta}^r$}};

\node [] (ar) at (3.86,0.92){\footnotesize{$\bm{\alpha}_v^r$}};

\node [] (br) at (3.47,0.43){\footnotesize{$\bm{\beta}^r$}};

\end{tikzpicture}

%% file: fastssc_tree_sscflip.tex
\begin{tikzpicture}[baseline = (v.center),
        level/.style={level distance = 6mm},
        level 1/.style={sibling distance=19mm, edge from parent/.style={draw,blue,line width=1.5pt}},
        level 2/.style={sibling distance=9.5mm, edge from parent/.style={draw,blue,line width=1pt}},
        level 3/.style={sibling distance=4.7mm, edge from parent/.style={draw,blue,line width=0.5pt}},
        ]

\tikzset{
frozen/.style={thick,draw=black,fill=white,minimum size=3mm,circle, inner sep=0},
fullspace/.style={thick,draw=black,fill=black,minimum size=3mm,circle, inner sep = 0},
mixed/.style={thick,draw=black,fill=gray!50,minimum size=3mm,circle, inner sep = 0},
phantom/.style={draw=white,fill=white,minimum size=3mm,circle, inner sep = 0},
birep/.style={thick,draw=black,pattern=north east lines,pattern color=purple,minimum size=3mm,circle, inner sep = 0},
spc/.style={thick,draw=black,pattern=crosshatch,pattern color=orange,minimum size=3mm,circle, inner sep = 0},
}

% circle around SPC node
\tikzset{green dotted/.style={draw=green!50!black, line width=1pt,
    dash pattern=on 3pt off 3pt,
    inner sep=0.4mm, rectangle, rounded corners}};

\node[mixed] (p){} [grow=left]
	child {node[birep] (2_0){}
	}
	child {node[spc] (v){} edge from parent[red]
	}
;

% circle around SPC node
%\node (g_concat) [green dotted, fit = (v)] {};

\node[rotate=-90] (spc_lab) at ($(v)+(0.4,0.0)$) {\small{\gls{spc}}};
\node[rotate=-90] (birep_lab) at ($(2_0)+(0.4,0.0)$) {\small{\gls{rep}}};

\end{tikzpicture}

%% file: sc_tree_sscflip.tex
\begin{tikzpicture}[baseline = (0_7.center),
        level/.style={level distance = 6mm},
        level 1/.style={sibling distance=19mm, edge from parent/.style={draw,blue,line width=1.5pt}},
        level 2/.style={sibling distance=9.5mm, edge from parent/.style={draw,blue,line width=1pt}},
        level 3/.style={sibling distance=4.7mm, edge from parent/.style={draw,blue,line width=0.5pt}},
        ]

\tikzset{
frozen/.style={thick,draw=black,fill=white,minimum size=3mm,circle, inner sep=0},
fullspace/.style={thick,draw=black,fill=black,minimum size=3mm,circle, inner sep = 0},
mixed/.style={thick,draw=black,fill=gray!50,minimum size=3mm,circle, inner sep = 0},
phantom/.style={draw=white,fill=white,minimum size=3mm,circle, inner sep = 0},
}

\tikzset{
parallel segment/.style={
   segment distance/.store in=\segDistance,
   segment pos/.store in=\segPos,
   segment length/.store in=\segLength,
   to path={
   ($(\tikztostart)!\segPos!(\tikztotarget)!\segLength/2!(\tikztostart)!\segDistance!90:(\tikztotarget)$) -- 
   ($(\tikztostart)!\segPos!(\tikztotarget)!\segLength/2!(\tikztotarget)!\segDistance!-90:(\tikztostart)$)  \tikztonodes
   }, 
   % Default values
   segment pos=.5,
   segment length=2.5ex,
   segment distance=-1mm,
},
}

\tikzset{green dotted/.style={draw=green!50!black, line width=0.75pt,
    dash pattern=on 3pt off 3pt,
    inner sep=0.4mm, rectangle, rounded corners}};

\node[mixed] (p){} [grow=left]
	child {node[mixed] (2_0){}
		child {node[frozen,dashed ] (1_0){}
		}
		child {node[mixed] (1_2){} edge from parent[red]
			child {node[fullspace, fill=white,dashed] (0_2){}
			}
			child {node[fullspace] (0_3){} edge from parent[red]
			}
		}
	}
	child {node[mixed] (v){\rotatebox{-90}{}} edge from parent[red]
		child {node[mixed] (cl){}
			child {node[frozen] (0_4){}
			}
			child {node[fullspace] (0_5){} edge from parent[red]
			}
		}
		child {node[mixed] (cr){} edge from parent[red]
			child {node[fullspace] (0_6){}
			}
			child {node[fullspace] (0_7){} edge from parent[red]
        % child {node[phantom] (p0_7){} edge from parent[white]
        % }
			}
		}
	}
;

\end{tikzpicture}

%% file: Flip_Distrib_dscf3_r025.tex
\begin{tikzpicture}
  \pgfplotsset{
    label style = {font=\fontsize{9pt}{7.2}\selectfont},
    tick label style = {font=\fontsize{9pt}{7.2}\selectfont}
  }
   
   \begin{axis}[%
    width=\textwidth,
    height=\plotfigureheight\columnwidth,
    xmin=0, xmax=266,
    xtick={0,40,...,280},
    xlabel={Index of the information bit, $j$},
    xlabel style={yshift=0.4em},
    ymin=0, ymax=0.08,
    ylabel style={yshift=-0.2em},
    ylabel={PMF, $\mathbb{P}(i_1=a_j)$},
    xlabel style={yshift=-0.2em},
    yminorticks, xmajorgrids,
    ymajorgrids, yminorgrids,
    legend pos=north west,
    mark size=1.3pt,
    line width=0.8pt
    ] 

 \addplot[ybar, bar width=0.1, fill=MyBlue, draw=MyBlue]table[x=bfidx,y=distr]{Distr_dscf3_N1024_5G_r025};

\addplot [color=red, dashed, line width=1.2] coordinates {(38, 0.08) (38, 0)}; 

\node[rotate=90] at (axis cs:44,0.05){$j_{\text{RHS}}=38$};  

\node[draw,fill=white] at (axis cs:170,0.065) {$\displaystyle \mathbb{P}_{\text{LHS}}=\sum_{j=0}^{37} \mathbb{P}(i_1=a_j)=0.59$};
\node[draw,fill=white] at (axis cs:170,0.03) {$\displaystyle \mathbb{P}_{\text{RHS}}=\sum_{j=38}^{266} \mathbb{P}(i_1=a_j)=0.41$};
\end{axis}

\end{tikzpicture}%

%% file: Saved_tree_branches.tex
\begin{tikzpicture}
% AC of N=16 : 1111111010000000

\node [draw, fill=white, circle, line width=0.8pt, outer sep=0pt, inner sep=0.12 cm, label=south:{$\hat{u}_0$}] (u0) at (0,0){};
\node [draw, fill=white, circle, line width=0.8pt, outer sep=0pt, inner sep=0.12 cm, label=south:{$\hat{u}_1$}] (u1) at (0.8,0){};
\node [draw, fill=white, circle, line width=0.8pt, outer sep=0pt, inner sep=0.12 cm, label=south:{$\hat{u}_2$}] (u2) at (1.6,0){};
\node [draw,  fill=white, circle, line width=0.8pt, outer sep=0pt, inner sep=0.12 cm, label=south:{$\hat{u}_3$}] (u3) at (2.4,0){};
\node [draw,  fill=white, circle, line width=0.8pt, outer sep=0pt, inner sep=0.12 cm, label=south:{$\hat{u}_4$}] (u4) at (3.2,0){};
\node [draw,  fill=white, circle, line width=0.8pt, outer sep=0pt, inner sep=0.12 cm, label=south:{$\hat{u}_5$}] (u5) at (4.0,0){};
\node [draw,  fill=black, circle, line width=0.8pt, outer sep=0pt, inner sep=0.12 cm, label=south:{$\hat{u}_6$}] (u6) at (4.8,0){};
\node [draw, fill=black, circle, line width=0.8pt, outer sep=0pt, inner sep=0.12 cm, label=south:{$\hat{u}_7$},] (u7) at (5.6,0){};

\node [draw, fill=white, circle, line width=0.8pt, outer sep=0pt, inner sep=0.12 cm, label=south:{$\hat{u}_8$}] (u8) at (6.4,0){};
\node [draw, fill=black,  circle, line width=0.8pt, outer sep=0pt, inner sep=0.12 cm, label=south:{$\hat{u}_9$}] (u9) at (7.2,0){};
\node [draw, circle, line width=0.8pt, outer sep=0pt, inner sep=0.12 cm, label=south:{$\hat{u}_{10}$}] (u10) at (8.0,0){};
\node [draw=red, fill=black, circle, line width=0.8pt, outer sep=0pt, inner sep=0.12 cm, label=south:{\textcolor{red}{$\hat{u}_{11}$}}] (u11) at (8.8,0){};
\node [draw, fill=black, circle, line width=0.8pt, outer sep=0pt, inner sep=0.12 cm, label=south:{$\hat{u}_{12}$}] (u12) at (9.6,0){};
\node [draw, fill=black, circle, line width=0.8pt, outer sep=0pt, inner sep=0.12 cm, label=south:{$\hat{u}_{13}$}] (u13) at (10.4,0){};
\node [draw, fill=black, circle, line width=0.8pt, outer sep=0pt, inner sep=0.12 cm, label=south:{$\hat{u}_{14}$}] (u14) at (11.2,0){};
\node [draw, fill=black, circle, line width=0.8pt, outer sep=0pt, inner sep=0.12 cm, label=south:{$\hat{u}_{15}$},] (u15) at (12.0,0){};

 \node [draw, circle, line width=0.8pt, outer sep=0pt, inner sep=0.12 cm, fill=gray!50!white] (n10) at ($(u0)+(0.4, 1.0)$){};
 \node [draw, circle, line width=0.8pt, outer sep=0pt, inner sep=0.12 cm, fill=gray!50!white] (n11) at ($(u2)+(0.4, 1.0)$){};
 \node [draw, circle, line width=0.8pt, outer sep=0pt, inner sep=0.12 cm, fill=gray!50!white] (n12) at ($(u4)+(0.4, 1.0)$){};
 \node [draw, circle, line width=0.8pt, outer sep=0pt, inner sep=0.12 cm, fill=gray!50!white] (n13) at ($(u6)+(0.4, 1.0)$){};

\node [draw, circle, line width=0.8pt, outer sep=0pt, inner sep=0.12 cm, fill=gray!50!white] (n14) at ($(u8)+(0.4, 1.0)$){};
 \node [draw, circle, line width=0.8pt, outer sep=0pt, inner sep=0.12 cm, fill=gray!50!white] (n15) at ($(u10)+(0.4, 1.0)$){};
 \node [draw, circle, line width=0.8pt, inner sep=0.12 cm, fill=gray!50!white] (n16) at ($(u12)+(0.4, 1.0)$){};
 \node [draw, circle, line width=0.8pt, inner sep=0.12 cm, fill=gray!50!white] (n17) at ($(u14)+(0.4, 1.0)$){};

 \node [draw, circle, line width=0.8pt, outer sep=0pt, inner sep=0.12 cm, fill=gray!50!white] (n20) at ($(n10)+(0.8, 1.0)$){};

 \node [draw, circle, line width=0.8pt, outer sep=0pt, inner sep=0.12 cm, fill=gray!50!white] (n21) at ($(n12)+(0.8, 1.0)$){};

 \node [draw, circle, line width=0.8pt, outer sep=0pt, inner sep=0.12 cm, fill=gray!50!white] (n22) at ($(n14)+(0.8, 1.0)$){};

 \node [draw, circle, line width=0.8pt, outer sep=0pt, inner sep=0.12 cm, fill=gray!50!white] (n23) at ($(n16)+(0.8, 1.0)$){};

 \node [draw, circle, line width=0.8pt, outer sep=0pt, inner sep=0.12 cm, fill=gray!50!white] (n30) at ($(n20)+(1.6, 1.0)$){};

 \node [draw, circle, line width=0.8pt, outer sep=0pt, inner sep=0.12 cm, fill=gray!50!white] (n31) at ($(n22)+(1.6, 1.0)$){};

 \node [draw, circle, line width=0.8pt, outer sep=0pt, inner sep=0.12 cm, fill=gray!50!white] (n40) at ($(n30)+(3.2, 1.0)$){};

 \draw (u0) -- (n10);
 \draw (u1) -- (n10);
 \draw (u2) -- (n11);
 \draw (u3) -- (n11);
 \draw (u4) -- (n12);
 \draw (u5) -- (n12);
 \draw (u6) -- (n13);
 \draw (u7) -- (n13);
 \draw (u8) -- (n14);
 \draw (u9) -- (n14);
 \draw (u10) -- (n15);
 \draw [color=red] (u11) -- (n15);
 \draw (u12) -- (n16);
 \draw (u13) -- (n16);
 \draw (u14) -- (n17);
 \draw (u15) -- (n17);

\draw [line width=0.8pt] (n10) -- (n20);
\draw [line width=0.8pt] (n11) -- (n20);
\draw [line width=0.8pt] (n12) -- (n21);
\draw [line width=0.8pt] (n13) -- (n21);

\draw [line width=0.8pt] (n14) -- (n22);
\draw [color=red, line width=0.8pt] (n15) -- (n22);
\draw (n16) -- (n23);
\draw [line width=0.8pt] (n17) -- (n23);

\draw [line width=1.2pt] (n20) -- (n30);
\draw [line width=1.2pt] (n21) -- (n30);
\draw [color=red, line width=1.2pt]  (n22) -- (n31);
\draw (n23) -- (n31);

\draw [line width=1.2pt] (n30) -- (n40);
\draw [color=red, line width=1.2pt] (n31) -- (n40);

%% draw stage lines and their connections
\node[inner sep=2] (s4) at ($(n40)+(-6.5,0)$) {$4$};
\node[inner sep=2] (s3) at ($(n30)+(-3.3,0)$) {$3$};
\node[inner sep=2] (s2) at ($(n20)+(-1.7,0)$){$2$};
\node[inner sep=2] (s1) at ($(n10)+(-0.9,0)$){$1$};
\node[inner sep=2] (s0) at ($(u0)+(-0.5,0)$){$0$};
% label s
\node[] (s_lab) at ($(s4)+(0.0,0.6)$) {$s\left(\varsigma\right)$};
%\draw ($(s4)+(-0.1,0.55)$)--($(s4)+(0.1,0.55)$);
%\draw ($(s_lab)+(-0.1,-0.12)$)--($(s_lab)+(0.1,-0.12)$);
%\node[] at ($(s4)+(0.0,0.8)$) {$_$};

\draw[dotted] (s4)--(n40);
\draw[dotted] (s3)--(n30)--(n31);
\draw[dotted] (s2)--(n20)--(n21)--(n22)--(n23);
\draw[dotted] (s1)--(n10)--(n11)--(n12)--(n13)--(n14)--(n15)--(n16)--(n17);
\draw[dotted] (s0)--(u0)--(u1)--(u2)--(u3)--(u4)--(u5)--(u6)--(u7)--(u8)--(u9)--(u10)--(u11)--(u12)--(u13)--(u14)--(u15);

%% skipped area
\draw[dashed, color=MyBlue, line width=1.2pt] ($(u10)+(0.5,-0.8)$) -- ($(u0)+(-0.5,-0.8)$) -- ($(n20)+(-0.7,0.4)$) -- ($(n30)+(1.0,1.0)$) -- ($(n22)+(-1.0,-0.4)$) -- ($(u10)+(0.5,0.2)$) -- ($(u10)+(0.5,-0.8)$);
%-- ($(u9)+(1.0,-0.4)$);

% Labels of saved and restart paths
% if needed add marker: -- node[]{$\bm{\times}$} 
\node[] (rest_lab) at (8.7,4.2) {\textcolor{red}{restart path}};
\draw[-latex,color=red] ($(rest_lab)+(0.1,-0.2)$) -- ($(rest_lab)+(-0.2,-0.68)$);

\draw[-latex, color=red] ($(n40)+(1.0,-0.1)$) -- node[above]{$\bm{\alpha}_3$} ($(n40)+(2.0,-0.4)$);

%--  -- node{} ($(rest_lab)+(0.5,-0.9)$);
% $\bm{\times}$
% lines over the restart path
\draw[-latex, color=red] ($(n31)+(-0.7,-0.2)$) -- node[yshift=0.2cm,xshift=-0.1cm]{$\bm{\alpha}_2$}($(n31)+(-1.27,-0.55)$);
\draw[-latex, color=red] ($(n22)+(0.4,-0.1)$) -- node[yshift=0.1cm,xshift=0.2cm]{$\bm{\alpha}_1$} ($(n22)+(0.8,-0.6)$);
\draw[-latex, color=red] ($(n15)+(0.25,-0.2)$) -- node[yshift=0.1cm,xshift=0.35cm]{$\alpha_{0}$} ($(n15)+(0.5,-0.7)$);

\node[inner sep=0.05cm] (alpha_ch) at ($(n40)+(0.0,0.7)$){$\bm{\alpha}_{\text{ch}}$};
\draw[-latex] (alpha_ch) -- (n40);

\node[] (skipped_lab) at (2.8,4.3) {\textcolor{MyBlue}{skipped nodes}};
\draw[-latex, color=MyBlue] ($(skipped_lab)+(0.1,-0.2)$) -- ($(skipped_lab)+(0.33,-0.64)$);
%-- ($(skipped_lab)+(-0.01,-0.65)$) -- ($(skipped_lab)+(-0.5,-1.0)$);

\node[draw=MyDarkGreen, rectangle, minimum width=0.6cm, minimum height=1.0cm, line width=1.2pt] at ($(u9)+(0.0,-0.2)$){};

%\node[inner sep=0.05cm] (beta_4) at ($(n40)+(0.0,0.7)$){$\bm{\beta}_{\text{3}}$};

% beta restoration
\draw[-latex, color=red] ($(n30)+(2.4,0.6)$) -- node[yshift=-0.25cm,xshift=0.04cm]{$\bm{\beta}_3$} ($(n30)+(3.0,0.8)$);

\draw[-latex, color=red] ($(n14)+(0.5,0.4)$) -- node[yshift=-0.20cm,xshift=0.15cm]{$\bm{\beta}_1$} ($(n14)+(0.8,0.8)$);

\draw[-latex] (alpha_ch) -- (n40);

\end{tikzpicture}

%% file: Funct_save_GRM_P.tex
\begin{tikzpicture}
  \pgfplotsset{
    label style = {font=\fontsize{9pt}{7.2}\selectfont},
    tick label style = {font=\fontsize{9pt}{7.2}\selectfont}
  }
   
   \begin{axis}[%
    width=\columnwidth,
    height=\plotfigureheight\columnwidth,
    xmin=0, xmax=522,
    xlabel={Index of the information bit, $j$},
    xlabel style={yshift=0.4em},
    ymin=0, ymax=3.5e3,
    ytick={0,500,...,3500},
    ylabel style={yshift=-0.2em},
    ylabel={Exec.-time reduction, CCs},
    xlabel style={yshift=-0.2em},
    yminorticks, xmajorgrids,
    ymajorgrids, yminorgrids,
    %legend pos=south east,
    legend style={nodes={scale=0.8}},
    legend style={at={(1.0,0.0)}, anchor=south east, legend cell align=left},    
    legend style={font=\small, column sep=0.3mm, row sep=-0.5mm},
    mark size=1.6pt, line width=0.8pt, mark options=solid,    
    ] 

 \addplot [mark=x, color=MyDarkGreen, mark repeat=80, mark phase=0]
 table[x=bfidx,y=CyclesSavedP64]{Saved_cycles_N1024_P.tex};
 \addlegendentry{\gls{grm} $P=64$ \eqref{eq:tot_saved_cc}} 
 
 \addplot [mark=square, color=MyBlue, mark repeat=80, mark phase=0]
 table[x=bfidx,y=CyclesSavedP16]{Saved_cycles_N1024_P.tex};
 \addlegendentry{\gls{grm} $P=16$ \eqref{eq:tot_saved_cc}}

  \addplot [color=MyOrange] coordinates {(0, 3099) (527, 3099)}; 
  \addlegendentry{\gls{sc} ref. $P=64$ \eqref{eq:lat_sc}} 
  
 \addplot [dashed, color=MyOrange] coordinates {(0, 3389) (527, 3389)}; 
  \addlegendentry{\gls{sc} ref. $P=16$ \eqref{eq:lat_sc}}

\end{axis}    
\end{tikzpicture}%

%% file: Funct_save_GRM_SRM.tex
\begin{tikzpicture}
  \pgfplotsset{
    label style = {font=\fontsize{9pt}{7.2}\selectfont},
    tick label style = {font=\fontsize{9pt}{7.2}\selectfont}
  }
   
   \begin{axis}[%
    width=\columnwidth,
    height=\plotfigureheight\columnwidth,
    xmin=0, xmax=522,
    xlabel={Index of the information bit, $j$},
    xlabel style={yshift=0.4em},
    ymin=0, ymax=3.5e3,
    ytick={0,500,...,3500},
    ylabel style={yshift=-0.2em},
    ylabel={Exec.-time reduction, CCs},
    xlabel style={yshift=-0.2em},
    yminorticks, xmajorgrids,
    ymajorgrids, yminorgrids,
   % legend pos=south east,
    legend style={nodes={scale=0.8}},    
    legend style={legend columns=2},   
    legend style={at={(1.0,0.15)}, anchor=south east,font=\small}, 
    legend style={column sep=0.1mm, row sep=-0.1mm, inner sep=0.4mm, legend cell align=left},
    mark size=1.6pt, line width=0.8pt, mark options=solid,
    ] 

    % font=\scriptsize
%  legend style={at={(0.005,0.65)}, 
 % \addplot [mark=triangle, color=MyBlue, mark repeat=50, mark phase=50]
 % table[x=bfidx,y=CyclesSavedP16]{data/Saved_cycles_N1024_P.tex};
 % \addlegendentry{Saved \gls{cc} \eqref{eq:l_p_saved_calc} $P=16$} 
    
 \addplot [mark=x, color=MyDarkGreen, mark repeat=80, mark phase=0]
 table[x=bfidx,y=CyclesSavedP64]{Saved_cycles_N1024_P.tex};
 \addlegendentry{\gls{grm} \eqref{eq:tot_saved_cc}} 

 \addplot [mark=triangle, color=red, mark repeat=80, mark phase=0]
 table[x=bfidx,y=CyclesSavedP64]{Saved_cycles_N1024_SRM.tex};
 \addlegendentry{\gls{srm} \cite{simp_rest_mech}} 

 \addplot [mark=*, color=black, mark repeat=80, mark phase=0]
 table[x=bfidx,y=CyclesSavedP64]{Saved_cycles_N1024_First_Info.tex};
 \addlegendentry{LRT \cite{Giard_JETCAS_2017}} 

  \addplot [color=MyOrange] coordinates {(0, 3099) (527, 3099)}; 
  \addlegendentry{\gls{sc} ref. \eqref{eq:lat_sc}} 

\end{axis}    
\end{tikzpicture}%

%% file: Mem_struc.tex
\begin{tikzpicture}[]

%% SC
\node [draw, fill=yellow!20, rectangle, align=center, inner sep=3pt,
minimum height=0.7cm, minimum width=1.3cm] (llr_ch) at (0.3,0) {$\bm{\alpha}_{\text{ch}}$}; 

\node [draw, fill=yellow!20, rectangle, align=center, inner sep=3pt,
minimum height=0.7cm, minimum width=1.3cm] (llr_int) at (0.3,-1.5) {$\bm{\alpha}_{\text{int}}$}; 

\node [draw, rectangle, fill=yellow!20, align=center, inner sep=3pt, minimum height=0.7cm, minimum width=1.3cm] (ps) at (2.4,0.0) {$\bm{\hat{u}}$}; 

\node [draw, rectangle, fill=yellow!20, align=center, inner sep=3pt, minimum height=0.7cm, minimum width=1.3cm] (uhat) at (2.4,-1.5) {$\bm{\beta}_{\text{int}}$}; 

\node [draw, fill=MyBlue!30, rectangle, align=center, inner sep=3pt, minimum height=0.7cm, minimum width=1.3cm] (metr_flip) at ($(ps.east)+(1.6,0)$) {$\bm{\mathcal{M}}_{\text{flip}}$}; 
\node [draw, fill=MyBlue!30, rectangle, align=center, inner sep=3pt, minimum height=0.7cm, minimum width=1.3cm] (bit_flip) at ($(uhat.east)+(1.6,0)$) {$\bm{\mathcal{B}}_{\text{flip}}$};

% \node [draw, fill=red!20, rectangle, align=center, inner sep=3pt, minimum height=0.7cm, minimum width=1.3cm] (ps_rest) at ($(metr_flip.east)+(1.6,0)$) {$\bm{\beta}_{\text{rest}}$}; 
\node [draw, fill=MyDarkGreen!30, rectangle, align=center, inner sep=3pt, minimum height=0.7cm, minimum width=1.3cm] (uhat_rest) at ($(bit_flip.east)+(1.5,0.8)$) {$\bm{\hat{u}}_{\text{rest}}$}; 

%% Memory lengths
%%% SC
\draw [<->] ($(llr_ch.south)+(-0.6,-0.2)$) -- node[below,font=\footnotesize]{$Q_{\text{ch}}$} ($(llr_ch.south)+(0.6,-0.2)$);
\draw [<->] ($(llr_ch.west)+(-0.2,-0.35)$) -- node[above,rotate=90,font=\footnotesize]{$N$}($(llr_ch.west)+(-0.2,0.35)$);

\draw [<->] ($(llr_int.west)+(-0.2,-0.35)$) -- node[above,rotate=90,font=\footnotesize]{$N-1$}($(llr_int.west)+(-0.2,0.35)$);
\draw [<->] ($(llr_int.south)+(-0.6,-0.2)$) -- node[below,font=\footnotesize]{$Q_{\text{int}}$} ($(llr_int.south)+(0.6,-0.2)$);

\draw [<->] ($(ps.south)+(-0.6,-0.2)$) -- node[below,font=\footnotesize]{$1$} ($(ps.south)+(0.6,-0.2)$);
\draw [<->] ($(ps.west)+(-0.2,-0.35)$) -- node[above,rotate=90,font=\footnotesize]{$N-1$}($(ps.west)+(-0.2,0.35)$);

\draw [<->] ($(uhat.south)+(-0.6,-0.2)$) -- node[below,font=\footnotesize]{$1$} ($(uhat.south)+(0.6,-0.2)$);
\draw [<->] ($(uhat.west)+(-0.2,-0.35)$) -- node[above,rotate=90,font=\footnotesize]{$N$}($(uhat.west)+(-0.2,0.35)$);

%%% SCF
\draw [<->] ($(metr_flip.south)+(-0.6,-0.2)$) -- node[below,font=\footnotesize]{$Q_{\text{flip}}$}($(metr_flip.south)+(0.6,-0.2)$);
\draw [<->] ($(metr_flip.west)+(-0.2,-0.35)$) -- node[above,rotate=90,font=\footnotesize]{$T_{\text{max}}-1$} ($(metr_flip.west)+(-0.2,0.35)$);

\draw [<->] ($(bit_flip.south)+(-0.6,-0.2)$) -- node[below,font=\footnotesize]{$\omega \times n$} ($(bit_flip.south)+(0.6,-0.2)$);
\draw [<->] ($(bit_flip.west)+(-0.2,-0.38)$) -- node[above,rotate=90,font=\footnotesize]{$T_{\text{max}}-1$}($(bit_flip.west)+(-0.2,0.38)$);

%%% SCF SRM
%\draw [<->] ($(ps_rest.south)+(-0.64,-0.2)$) -- node[below]{$1$} ($(ps_rest.south)+(0.65,-0.2)$);
%\draw [<->] ($(ps_rest.west)+(-0.2,-0.35)$) -- node[above,rotate=90,font=\footnotesize]{$\nicefrac{N}{2}$}($(ps_rest.west)+(-0.2,0.35)$);

\draw [<->] ($(uhat_rest.south)+(-0.65,-0.2)$) -- node[below]{$1$} ($(uhat_rest.south)+(0.65,-0.2)$);
\draw [<->] ($(uhat_rest.west)+(-0.2,-0.35)$) -- node[above,rotate=90,font=\footnotesize]{$N$}($(uhat_rest.west)+(-0.2,0.35)$);

%% Dashed borders
%% SC
\node [font=\footnotesize] (sc_lab) at ($(llr_ch.north) + (-1.1,0.52)$) {\gls{sc}}; 

\draw[dashed] ($(llr_ch.north) + (-1.3,0.3)$) -| ($(uhat.south east)+(0.2,-0.7)$) -| ($(llr_ch.north) + (-1.3,0.3)$);

%SCF
\node [font=\footnotesize, anchor=west] (scf_lab) at ($(sc_lab.west) + (0,0.4)$) {\gls{scf}};

\draw[dashed] ($(llr_ch.north) + (-1.5,0.75)$) -| ($(bit_flip.south east)+(0.2,-0.9)$) -|  ($(llr_ch.north) + (-1.5,0.75)$);

% SCF-GRM
\node [font=\footnotesize, anchor=west] (scf_srm_lab) at ($(scf_lab.west) + (0,0.35)$) {\gls{scf} with  \gls{grm}}; 

%\draw[dashed] ($(llr_ch.north) + (-1.9,1.1)$) -- ($(llr_ch.north)+(5.8,1.1)$) -- ($(llr_ch.north)+(5.8,-4.8)$) -- ($(metr_flip.south) + (-1.9,-0.9)$) -- ($(llr_ch.north) + (-1.9,1.1)$);
\draw[dashed] ($(llr_ch.north) + (-1.7,1.1)$) -| ($(uhat_rest.south east)+(0.2,-1.9)$) -| ($(llr_ch.north) + (-1.7,1.1)$);

\end{tikzpicture}

%% file: FER_SCF_SCL.tex
\begin{tikzpicture}
  \pgfplotsset{
    label style = {font=\fontsize{9pt}{7.2}\selectfont},
    tick label style = {font=\fontsize{9pt}{7.2}\selectfont}
  }

  \begin{semilogyaxis}[%
    width=0.975*\columnwidth,
    height=0.975*\plotfigureheight\columnwidth,
    xmin=0.5, xmax=2.75,
    xlabel={$\nicefrac{E_b}{N_0},\,\mathrm{dB}$},
    xlabel style={yshift=0.4em},
    ymin=1e-4, ymax=1,
    ylabel style={yshift=-0.2em},
    ylabel={Frame-error rate},
    yminorticks, xmajorgrids,
    ymajorgrids, yminorgrids,
    legend pos=south west,
    legend style={nodes={scale=0.9}},
    legend style={at={(0.0,0.0)}, anchor=south west},   
    legend style={legend columns=2, font=\footnotesize, column sep=0mm, row sep=-0.5mm,  legend cell align=left},
    mark size=1.8pt, line width=0.8pt, mark options=solid,
    ]  

 \addplot [color=MyOrange, mark=pentagon]
  table[x=EbN0,y=fer]{scf_exec_av_N1024_5G_R05.tex};
    \addlegendentry{\gls{scf}}
    
    \addplot [color=red, mark=triangle]
  table[x=EbN0,y=fer]{dscf1_exec_av_N1024_5G_R05.tex};
    \addlegendentry{\gls{dscf}-1}

    \addplot [color=MyBlue, mark=diamond]
  table[x=EbN0,y=fer]{dscf2_exec_av_N1024_5G_R05.tex};
    \addlegendentry{\gls{dscf}-2}

  %   \addplot [color=MyDarkGreen, mark=diamond]
  % table[x=EbN0,y=fer]{data/Exec_Time_Sav/dscf3/dscf3_exec_av_N1024_R05.tex};
  %   \addlegendentry{\gls{dscf}-3} 

    \addplot [color=MyDarkGreen, mark=square]
  table[x=EbN0,y=fer]{dscf3_exec_av_N1024_5G_R05.tex};
    \addlegendentry{\gls{dscf}-3} % also 5G

   \addplot [color=black, mark=x]
  table[x=EbN0,y=fer]{scl_l8_N1024_R05_5G.tex};
    \addlegendentry{\gls{scl} $L=8$} 

  %   \addplot [color=black, dashed, mark=x]
  % table[x=EbN0,y=fer]{data/scl/scl_l16_N1024_R05_5G.tex};
  %   \addlegendentry{\gls{scl} $L=16$} 

  \end{semilogyaxis}

\end{tikzpicture}%

%% file: FER_DSCFw.tex
\begin{tikzpicture}
  \pgfplotsset{
    label style = {font=\fontsize{9pt}{7.2}\selectfont},
    tick label style = {font=\fontsize{9pt}{7.2}\selectfont}
  }

  \begin{semilogyaxis}[%
    width=0.975*\columnwidth,
    height=0.975*\plotfigureheight\columnwidth,
    xmin=-6.0, xmax=2.5,
    xlabel={SNR,$\,\mathrm{dB}$},
    xlabel style={yshift=0.4em},
    ymin=1e-4, ymax=1.2,
    ylabel style={yshift=-0.2em},
    ylabel={Frame-error rate},
    yminorticks, xmajorgrids,
    ymajorgrids, yminorgrids,
    %legend pos=south west,
    legend style={at={(0.02,0.02)},anchor=south west},
    legend style={nodes={scale=0.9}},
    legend style={at={(0.0,0.75)}, anchor=south west},   
    legend style={legend columns=2, font=\footnotesize, column sep=0mm, row sep=-0.7mm, inner sep=0.11mm},
    legend style={fill=white, fill opacity=0.75, draw opacity=1,text opacity=1},
    mark size=1.8pt, line width=0.8pt, mark options=solid,
    ]  

   \addplot [color=red]
  table[x=SNR,y=fer]{dscf3_exec_av_N1024_5G_R0125.tex};
    \addlegendentry{\gls{dscf}-3 $R=\nicefrac{1}{8}$}
    
    \addplot [color=red, mark=square, mark repeat=2]
  table[x=SNR,y=fer]{dscf3_exec_av_N1024_5G_R0125.tex};
    \addlegendentry{w. \gls{grm}}

    \addplot [color=MyBlue]
  table[x=SNR,y=fer]{dscf3_exec_av_N1024_5G_R025.tex};
    \addlegendentry{\gls{dscf}-3 $R=\nicefrac{1}{4}$}
    
    \addplot [color=MyBlue, mark=triangle, mark repeat=2]
  table[x=SNR,y=fer]{dscf3_exec_av_N1024_5G_R025.tex};
    \addlegendentry{w. \gls{grm}}                    

    \addplot [color=MyDarkGreen]
  table[x=EbN0,y=fer]{dscf3_exec_av_N1024_5G_R05.tex};
    \addlegendentry{\gls{dscf}-3 $R=\nicefrac{1}{2}$} 

   \addplot [color=MyDarkGreen, mark=diamond, mark repeat=2]
  table[x=EbN0,y=fer]{dscf3_exec_av_N1024_5G_R05.tex};
    \addlegendentry{w. \gls{grm}} 
    
  \end{semilogyaxis}

\end{tikzpicture}%

%% file: Aver_exec_time_GRM_SRM_R0125.tex
\begin{tikzpicture}
  \pgfplotsset{
    label style = {font=\fontsize{9pt}{7.2}\selectfont},
    tick label style = {font=\fontsize{9pt}{7.2}\selectfont}
  }

   \begin{semilogyaxis}[%
    width=\columnwidth,
    height=\plotfigureheight\columnwidth,
    xmin=3.2e-04, xmax=1.05e-01,
    xlabel={Frame-error rate},
    xlabel style={yshift=0.4em},
    ymin=2e3, ymax=2e5,
    x dir=reverse,
    xmode=log,
    ylabel style={yshift=-0.6em},
    ylabel={Avg. Exec. Time, CCs},
    xlabel style={yshift=-0.2em},
    yminorticks, xmajorgrids,
    ymajorgrids, yminorgrids,
    legend style={at={(1.0,1.0)},anchor=north east},
    legend style={legend columns=2, font=\footnotesize, column sep=0.5mm, row sep=-0.5mm, legend cell align=left}, 
    mark size=1.8pt, mark options=solid,
    ] 
      
  \addplot [line width=0.8pt, mark=o, color=black]
  table[x=fer,y=execav]{dscf3_exec_av_N1024_5G_R0125.tex};
  \addlegendentry{\gls{dscf}-3}

  \addplot [line width=0.8pt, mark=triangle, color=red]
  table[x=fer,y=execavsrm]{dscf3_exec_av_N1024_5G_R0125.tex};
  \addlegendentry{w. \gls{srm}}

  %   \addplot [line width=0.8pt, mark=diamond, color=MyBlue]
  % table[x=fer,y=execavplrt]{data/Exec_Time_Sav/dscf3/dscf3_exec_av_N1024_5G_R0125.tex};
  % \addlegendentry{w. $\text{PLRT}^{*}$}

  %   \addplot [line width=0.8pt, mark=diamond, color=MyBlue]
  % table[x=fer,y=execavplrt]{data/Exec_Time_Sav/dscf3/dscf3_PLRT_exec_av_N1024_5G_R0125.tex};
  % \addlegendentry{w. PolarBear}
 % \addlegendentry{w. $\text{PLRT}$}
  
  \addplot [line width=0.8pt, mark=square, color=MyDarkGreen]
  table[x=fer,y=execavgrm]{dscf3_exec_av_N1024_5G_R0125.tex};
  \addlegendentry{w. \gls{grm}}
    
  \addplot[color=MyOrange,  mark=none, line width=0.8pt] table[row sep=crcr]{%
    1 3099\\    
    1e-1 3099\\    
    1e-2 3099\\    
    1e-3 3099\\    
    3e-4 3099\\    
    };
  \addlegendentry{\gls{sc} ref. \eqref{eq:lat_sc}}

\node[align=center] at (axis cs:0.04,4100) {$ \displaystyle\mathcal{L_{\text{SC}}}=3099$};

\end{semilogyaxis}    
\end{tikzpicture}%

%% file: Aver_exec_time_GRM_SRM_R025.tex
\begin{tikzpicture}

  \pgfplotsset{
    label style = {font=\fontsize{9pt}{7.2}\selectfont},
    tick label style = {font=\fontsize{9pt}{7.2}\selectfont}
  }
   \begin{semilogyaxis}[%
    width=\columnwidth,
    height=\plotfigureheight\columnwidth,
    xmin=3.2e-04, xmax=1.05e-01,
    xlabel={Frame-error rate},
    xlabel style={yshift=0.4em},
    ymin=2e3, ymax=2e5,
    x dir=reverse,
    xmode=log,
    ylabel style={yshift=-0.6em},
    ylabel={Avg. Exec. Time, CCs},
    xlabel style={yshift=-0.2em},
    yminorticks, xmajorgrids,
    ymajorgrids, yminorgrids,
    legend style={at={(1.0,1.0)},anchor=north east},
    legend style={legend columns=2, font=\footnotesize, column sep=0.5mm, row sep=-0.5mm, legend cell align=left}, 
    mark size=1.8pt, mark options=solid,
    ] 
      
  \addplot [line width=0.8pt, mark=o, color=black]
  table[x=fer,y=execav]{dscf3_exec_av_N1024_5G_R025.tex};
  \addlegendentry{\gls{dscf}-3}

  \addplot [line width=0.8pt, mark=triangle, color=red]
  table[x=fer,y=execavsrm]{dscf3_exec_av_N1024_5G_R025.tex};
  \addlegendentry{w. \gls{srm}}

  % \addplot [line width=0.8pt, mark=diamond, color=MyBlue]
  % table[x=fer,y=execavplrt]{data/Exec_Time_Sav/dscf3/dscf3_exec_av_N1024_5G_R025.tex};
  % \addlegendentry{w. $\text{PLRT}^{*}$}

  % \addplot [line width=0.8pt, mark=diamond, color=MyBlue]
  % table[x=fer,y=execavplrt]{data/Exec_Time_Sav/dscf3/dscf3_PLRT_exec_av_N1024_5G_R025.tex};
  % \addlegendentry{w. PolarBear}
 % \addlegendentry{w. $\text{PLRT}$}

  \addplot [line width=0.8pt, mark=square, color=MyDarkGreen]
  table[x=fer,y=execavgrm]{dscf3_exec_av_N1024_5G_R025.tex};
  \addlegendentry{w. \gls{grm}}
    
  \addplot[color=MyOrange,  mark=none, line width=0.8pt] table[row sep=crcr]{%
    1 3099\\    
    1e-1 3099\\    
    1e-2 3099\\    
    1e-3 3099\\    
    3e-4 3099\\    
    };
  \addlegendentry{\gls{sc} ref. \eqref{eq:lat_sc}}

 \node[align=center] at (axis cs:0.04,4100) {$\displaystyle\mathcal{L_{\text{SC}}}=3099$};
% \draw[<->,red] (axis cs:0.01,19500) -- (axis cs:0.01,10800);

\end{semilogyaxis}    
\end{tikzpicture}%

%% file: Aver_exec_time_GRM_SRM_R05.tex
\begin{tikzpicture}

  \pgfplotsset{
    label style = {font=\fontsize{9pt}{7.2}\selectfont},
    tick label style = {font=\fontsize{9pt}{7.2}\selectfont}
  }

   \begin{semilogyaxis}[%
    width=\columnwidth,
    height=\plotfigureheight\columnwidth,
    xmin=3.2e-04, xmax=1.05e-01,
    xlabel={Frame-error rate},
    xlabel style={yshift=0.4em},
    ymin=2e3, ymax=2e5,
    x dir=reverse,
    xmode=log,
    ylabel style={yshift=-0.1em},
    ylabel={Avg. Exec. Time, CCs},
    xlabel style={yshift=-0.2em},
    yminorticks, xmajorgrids,
    ymajorgrids, yminorgrids,
    legend style={at={(1.0,1.0)},anchor=north east},
    legend style={legend columns=2, font=\footnotesize, column sep=0.5mm, row sep=-0.5mm, legend cell align=left}, 
    mark size=1.8pt, mark options=solid,
    ] 
      
  \addplot [line width=0.8pt, mark=o, color=black]
  table[x=fer,y=execav]{dscf3_exec_av_N1024_5G_R05.tex};
  \addlegendentry{\gls{dscf}-3}

  \addplot [line width=0.8pt, mark=triangle, color=red]
  table[x=fer,y=execavsrm]{dscf3_exec_av_N1024_5G_R05.tex};
  \addlegendentry{w. \gls{srm}}

  % \addplot [line width=0.8pt, mark=diamond, color=MyBlue]
  % table[x=fer,y=execavplrt]{data/Exec_Time_Sav/dscf3/dscf3_PLRT_exec_av_N1024_5G_R05.tex};
  % \addlegendentry{w. PolarBear}
 % \addlegendentry{w. $\text{PLRT}$}
  
  \addplot [line width=0.8pt, mark=square, color=MyDarkGreen]
  table[x=fer,y=execavgrm]{dscf3_exec_av_N1024_5G_R05.tex};
  \addlegendentry{w. \gls{grm}}
    
  \addplot[color=MyOrange,  mark=none, line width=0.8pt] table[row sep=crcr]{%
    1 3099\\    
    1e-1 3099\\    
    1e-2 3099\\    
    1e-3 3099\\    
    3e-4 3099\\    
    };
  \addlegendentry{\gls{sc} ref. \eqref{eq:lat_sc}}

 \node[align=center] at (axis cs:0.04,4100) {$ \displaystyle\mathcal{L_{\text{SC}}}=3099$};

\end{semilogyaxis}    
\end{tikzpicture}%

%% file: Aver_exec_time_all_R025.tex
\begin{tikzpicture}
  \pgfplotsset{
    label style = {font=\fontsize{9pt}{7.2}\selectfont},
    tick label style = {font=\fontsize{9pt}{7.2}\selectfont}
  }

   \begin{semilogyaxis}[%
    width=1.25\columnwidth,
    height=\plotfigureheight\columnwidth,
    xmin=3.2e-04, xmax=1.05e-01,
    xlabel={Frame-error rate},
    xlabel style={yshift=0.4em},
    ymin=500, ymax=1.5e5,
    x dir=reverse,
    xmode=log,
    ylabel style={yshift=-0.6em},
    ylabel={Avg. Exec. Time, CCs},
    xlabel style={yshift=-0.2em},
    yminorticks, xmajorgrids,
    ymajorgrids, yminorgrids,
    legend style={at={(1.0,1.0)},anchor=south east},
    legend style={legend columns=3, font=\footnotesize, column sep=0.41mm, row sep=-0.5mm, legend cell align=left}, 
    mark size=1.8pt, mark options=solid,
    ]      

  \addplot [line width=0.8pt, mark=o, color=black]
  table[x=fer,y=execav]{dscf3_exec_av_N1024_5G_R025.tex};
  \addlegendentry{\gls{dscf}-3}
  \addplot [line width=0.8pt, mark=square, color=MyDarkGreen]
  table[x=fer,y=execavgrm]{dscf3_exec_av_N1024_5G_R025.tex};
  \addlegendentry{w. \gls{grm}}  
  \addplot[color=MyOrange,  mark=none, line width=0.8pt] table[row sep=crcr]{%
    1 3099\\    
    1e-1 3099\\    
    1e-2 3099\\    
    1e-3 3099\\    
    3e-4 3099\\    
    };
  \addlegendentry{\gls{sc} ref. \eqref{eq:lat_sc}}
  
 \addplot [line width=0.8pt, mark=o, color=black,dashed]
  table[x=fer,y=execavplrt]{dscf3_PLRT_exec_av_N1024_5G_R025.tex};
  \addlegendentry{w. \gls{lrt}}
  
  \addplot [line width=0.8pt,mark=square, color=MyDarkGreen,dashed]
  table[x=fer,y=execavplrtgrm]{dscf3_PLRT_exec_av_N1024_5G_R025.tex};
  \addlegendentry{w. \gls{lrt}\,+\,\gls{grm}}
    
  \addplot[color=MyOrange, dashed,  mark=none, line width=0.8pt] table[row sep=crcr]{%
    1 2349\\    
    1e-1 2349\\    
    1e-2 2349\\    
    1e-3 2349\\    
    3e-4 2349\\    
    };
  \addlegendentry{$\scLRT$ ref. \eqref{eq:lat_scLRT}}

\node[align=center] at (axis cs:0.04,1600) {$ \displaystyle\mathcal{L}_{\scLRT}=2349$};

 \node[align=center] at (axis cs:0.045,4100) {$\displaystyle\mathcal{L_{\text{SC}}}=3099$};
  \addplot [line width=0.8pt, mark=o, color=black, dash dot]
  table[x=fer,y=execav]{dscf3_Fast_exec_av_N1024_5G_R025.tex};
  \addlegendentry{Fast-\gls{dscf}-3}  
  
  \addplot [line width=0.8pt, mark=square, color=MyDarkGreen,dash dot]
  table[x=fer,y=execavgrm]{dscf3_Fast_exec_av_N1024_5G_R025.tex};
  \addlegendentry{w. FAST\,+\,\gls{grm}}
    
  \addplot[color=MyOrange,  mark=none, line width=0.9pt, dash dot] table[row sep=crcr]{%
    1 741\\    
    1e-1 741\\    
    1e-2 741\\    
    1e-3 741\\    
    3e-4 741\\    
    };
  \addlegendentry{FSSC ref. \eqref{eq:fssc_lat}}

 \node[align=center] at (axis cs:0.04,1020) {$\displaystyle\mathcal{L_{\text{FSSC}}}=741$};

\end{semilogyaxis}    
\end{tikzpicture}%